\documentclass[12pt]{article}
\usepackage{amssymb,amsmath,amsthm,amsxtra,amsfonts}
\usepackage[margin=1in]{geometry}
\usepackage[onehalfspacing]{setspace}
\usepackage{setspace}
\usepackage{aliascnt}
\usepackage{color}
\usepackage[usenames,dvipsnames,table]{xcolor}
\usepackage{subcaption}
\usepackage{bbm}
\usepackage{dsfont}
\usepackage{graphicx}
\usepackage{mathrsfs}
\usepackage{tikz}
\usepackage{subcaption}
\usetikzlibrary{plotmarks,arrows,calc,patterns}

\usepackage{algorithm,algpseudocode}
\usepackage{threeparttable}
\usepackage{booktabs}
\usepackage{rotating}
\usepackage{lscape}
\usepackage{authblk}

\usepackage[round]{natbib}
\usepackage[colorlinks=true,citecolor=Blue,urlcolor=Blue,linkcolor=Blue]{hyperref}
\usepackage{setspace} 
\usepackage{booktabs} 
\usepackage{caption}
\usepackage{geometry} 
\usepackage{amsmath}
\usepackage{amsfonts}
\usepackage{array}
\usepackage{longtable}
\usepackage{xltabular}
\usepackage{booktabs}
\usepackage{enumitem}
\newcolumntype{L}[1]{>{\raggedright\arraybackslash}p{#1}}
\newcolumntype{C}[1]{>{\centering\arraybackslash}p{#1}}

\title{Words that Move Markets: Quantifying the Impact of RBI's Monetary Policy Communications on Indian Financial Market}
\author{
    Rohit Kumar\thanks{Indian Institute of Technology Delhi, Email: rohitk@iitd.ac.in},
    Sourabh Bikas Paul\thanks{Indian Institute of Technology Delhi, Email: sbpaul@hss.iitd.ac.in}\, and
        Nikita Singh\thanks{Indian Institute of Technology Delhi; University of Delhi, Email: nsingh9@db.du.ac.in}
}
\date{September 15, 2024}

\begin{document}

\maketitle

\begin{abstract}
We analyze the impact of the Reserve Bank of India's (RBI) monetary policy communications on Indian financial market from April 2014 to June 2024 using advanced natural language processing techniques. Employing BERTopic for topic modeling and a fine-tuned RoBERTa model for sentiment analysis, we assess how variations in sentiment across different economic topics affect the stock market. Our findings indicate that dovish sentiment generally leads to declines in equity markets, particularly in topics related to the interest rate policy framework and economic growth, suggesting that market participants interpret dovish language as signaling economic weakness rather than policy easing. Conversely, dovish sentiment regarding foreign exchange reserves management has a positive impact on equity market. These results highlight the importance of topic-specific communication strategies for central banks in emerging markets.

\end{abstract}



\section{Introduction}

Central bank communication has evolved into a critical instrument for steering monetary policy and shaping market expectations. As central banks worldwide embrace greater transparency, the volume and complexity of their communications have surged. This shift has created both opportunities and challenges for researchers, policymakers, and market participants seeking to understand and anticipate monetary policy decisions
and their impacts on financial markets.

The Reserve Bank of India (RBI) plays a pivotal role in guiding India's monetary policy. Its official mandate, as stated in its preamble, is:
``to regulate the issue of Bank notes and keeping of reserves with a view to securing monetary stability in India and generally to operate the currency and credit system of the country to its advantage; to have a modern monetary policy framework to meet the challenge of an increasingly complex economy, to maintain price stability while keeping in mind the objective of growth." \footnote{Accessed on 15 September 2024; \url{https://website.rbi.org.in/web/rbi/about-us}}

However, the dynamic nature of India's emerging economy often necessitates that the RBI navigate a complex landscape of economic challenges, at times requiring shifts in focus beyond its primary mandate. These subtle changes in policy emphasis are closely monitored and interpreted by market participants and analysts, who attempt to discern the broader implications for monetary policy direction.

\begin{figure}
    \centering
    \includegraphics[width=0.65\linewidth]{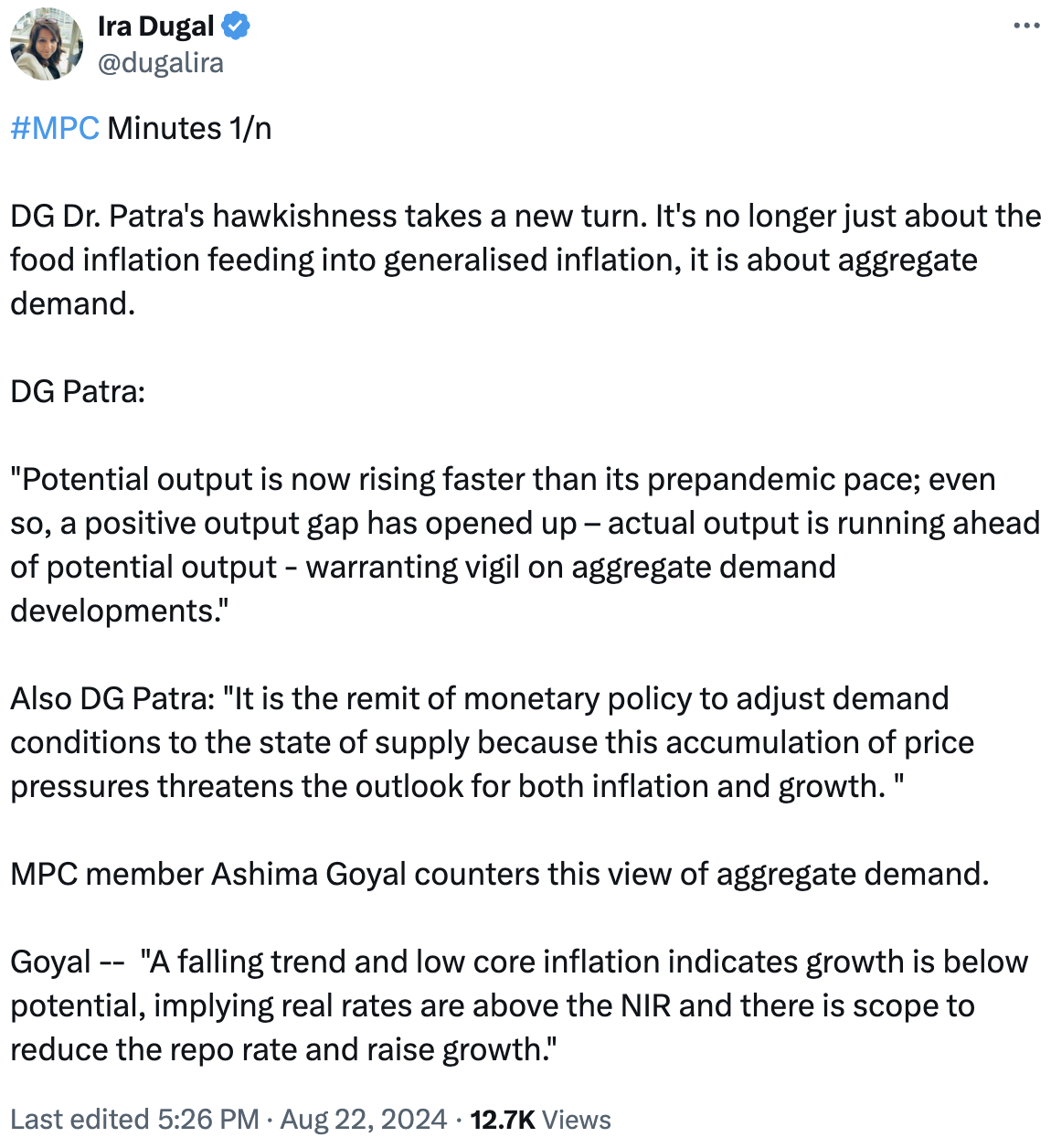}
    \caption{\textbf{Hawkish Signals in RBI Communications}
Financial journalist Ira Dugal provides insights into Deputy Governor Dr. Michael Patra’s increasingly hawkish stance in recent MPC discussions. The tweet highlights how subtle shifts in emphasis—such as moving from food inflation concerns to aggregate demand—can signal broader monetary policy shifts.}
    \label{fig:ira tweet}
\end{figure}

A recent discussion sparked by financial journalist Ira Dugal exemplifies this keen attention to nuances in RBI communications \footnote{Accessed on 15 September 2024; \url{https://x.com/dugalira/status/1826589190177652885}}. Dugal's analysis, focusing on the hawkish stance of Deputy Governor Dr. Michael Patra, underscores how subtle shifts in rhetoric and emphasis can have broader implications for interpreting the RBI's policy intentions (Figure~\ref{fig:ira tweet}). Such expert interpretations highlight the importance of systematic analysis of central bank communications to understand evolving policy priorities and their potential market impacts.

\newpage
In this paper, we aim to explore a decade of RBI monetary policy communications (April 2014- June 2024) through two central research questions:
\begin{enumerate}
    \item How does sentiment in the RBI’s Monetary Policy Committee (MPC) communications vary across different economic topics? 
    \item What does the variation in sentiment in RBI communications reveal about India’s monetary policy priorities and their potential impact on financial markets?
\end{enumerate}

This period is particularly significant as it marks the inception and establishment of bi-monthly monetary policy statements and regular press conferences by the RBI, representing a crucial shift towards more structured and frequent communication with the public and financial markets.

To address these questions, we employ state-of-the-art natural language processing (NLP) and machine learning techniques. Specifically, we utilize BERTopic to extract and classify key economic themes within RBI communications. A pre-trained RoBERTa model fine-tuned on central bank communication data (\cite{shah2023}) is employed to assess sentiment across these themes. By integrating sentiment analysis with topic modeling, we capture the nuances in RBI communications. Later on, we relate them to financial market movements through local projection methods to estimate the dynamic effects of RBI communications on different financial market indicators.

The combination of topic modeling and sentiment analysis is particularly effective for this study as it allows us to simultaneously identify key themes in RBI communications and quantify the associated sentiment. This dual approach captures both the content and tone of policy discussions, providing a more comprehensive view of the RBI's communication strategy and its potential market impacts.

This study offers several contributions. First, it provides a comprehensive, data-driven analysis of RBI communication strategies during a pivotal period in India’s economic development. Second, it quantifies the market impacts of RBI communications using intra-day financial data and local projection techniques. Third, our research bridges the application of cutting-edge NLP methodologies to the study of central bank communication within an emerging market context, an area where empirical research remains limited.

This research is motivated by the growing importance of the RBI and Indian financial
markets in the global economy. As an emerging market economy with unique challenges,
India provides an important case study for examining how central bank communication
strategies developed primarily in advanced economies may need to be adapted for different
contexts.

Our findings have important implications for improving the effectiveness of central bank communications in emerging markets. By quantifying how different aspects of RBI communications move markets, we can provide insights that can help policymakers optimize their messaging strategies. For investors and market participants, our results offer a framework for
extracting signals from central bank communications to inform trading and risk management
decisions.

The rest of the paper is organized as follows. Section 2 reviews the relevant literature. Section 3 details our data collection and preparation process. Section 4 discusses the methodology used for clustering and sentiment analysis. Section 5 presents data analysis and visualizations to explore RBI communication patterns. Section 6 outlines the empirical strategy, presents the empirical findings, and assesses their impact on Indian financial markets. Finally, Section 6 concludes with a discussion of policy implications.

\section{Literature Review}

Our paper builds on and contributes to several strands of literature on central bank communications, monetary policy in emerging markets, and applications of NLP to economic texts.

\textbf{Central Bank Communications.} A large body of research has examined how central bank communications influence financial markets and the macroeconomy. \cite{blinder2008} provide a comprehensive survey of this literature, emphasizing that greater transparency and communication have become central pillars of modern monetary policy frameworks. For major central banks like the Federal Reserve, ECB, and Bank of England, studies have shown that communications can significantly influence asset prices and shape inflation expectations (e.g., \cite{gurkaynak2005}; \cite{ehrmann2007}). 

Recent work has leveraged textual analysis to extract deeper insights from policy communications. \cite{hansen2016} apply computational linguistics to examine how the information content and tone of FOMC statements have evolved over time, particularly focusing on their role in signaling economic conditions. \cite{hansen2017} analyze FOMC transcripts to study dynamics of deliberation and consensus-building within the committee. They use text analysis to study how internal discussions influence decision-making and transparency. \cite{schmeling2019} provide evidence that the linguistic tone of central bank statements can move asset prices, underlining the importance of careful communication strategies.

Building on these insights, \cite{aruoba2024} proposed a natural language approach to identifying monetary policy shocks, further illustrating the power of NLP in uncovering nuanced information from central bank communications. Their approach complements traditional econometric methods, offering new avenues for interpreting central bank signals and their impact on financial markets.

While much of the existing literature has focused on central banks in advanced economies, there is a growing recognition of the importance of communication strategies in emerging markets. \cite{sil2023} applied text analysis techniques to the minutes of RBI’s Monetary Policy Committee meetings, identifying key themes and instances of dissent that influence monetary policy transmission. This work emphasizes the need to systematically examine the broader range of RBI communications over an extended period.

\textbf{Monetary Policy in Emerging Markets.} In emerging market economies like India, monetary policy frameworks face distinct challenges, including volatile capital flows and under-developed financial markets (\cite{mohanty2005}). For India, the transition to inflation targeting in 2016 marked a significant evolution in the RBI’s policy approach, reflecting a growing emphasis on communication and transparency (\cite{mohan2019}). This shift underscores the importance of analyzing the sentiment within the RBI’s Monetary Policy Committee (MPC) communications, as it offers insights into the priorities guiding these decisions and their implications for financial markets.

However, relatively few studies have conducted systematic analyses of RBI communications. \cite{mathur2019} examine the role of RBI communications in anchoring inflation expectations but focus primarily on a limited subset of policy statements within a narrow time-frame. In contrast, our work offers a more comprehensive analysis by covering a broader range of RBI communications and over an extended period, allowing for a deeper understanding of the evolving policy priorities and their implications.

\textbf{NLP Applications in Economics and Central Banking.} Advances in natural language processing (NLP) have facilitated novel analyses of economic and financial texts. \cite{gentzkow2019} provide an extensive overview of text-as-data methods in economics. They highlight applications from sentiment analysis to topic modeling. In the central banking context, \cite{gorodnichenko2023} analyze vocal patterns in central bank communications using advanced techniques, demonstrating how auditory cues—such as tone and stress—can offer additional insights into the information conveyed in policy announcements.

Further advancements have enabled more sophisticated analyses of central bank communications through text analysis and topic modeling techniques. \cite{ogunleye2023} compared various topic modeling approaches in the banking context, finding that transformer-based models like BERT outperform traditional methods such as Latent Dirichlet Allocation (LDA) in identifying coherent topics. This supports our choice of BERTopic for analyzing RBI communications. 

The use of advanced NLP techniques like BERTopic, as described by \cite{grootendorst2022}, \cite{devlin2019}, offers a powerful tool for uncovering thematic structures in large corpora of text. Our study applies these methods to track the evolution of key economic themes in RBI communications over a decade, providing insights into how the central bank’s priorities have shifted in response to different economic conditions.

\cite{born2013} demonstrated the potential of social media data in monetary policy analysis, using Twitter to construct real-time measures of inflation expectations. While our study focuses on official RBI communications, their work highlights the growing importance of diverse text sources in understanding monetary policy dynamics. \cite{ruman2023} conducted a comparative textual analysis of FOMC transcripts during inflation peaks, revealing how central bank discussions evolve during periods of economic stress. This underscores the value of examining RBI communications across different economic conditions to uncover shifts in policy priorities and communication strategies.

\textbf{Local Projections Method} Recent work by \cite{plagborg2021local} has shown that local projections (LPs) and Vector Autoregressions (VARs) estimate the same impulse responses in population, under fairly general conditions. However, LPs offer several practical advantages that make them particularly suitable for our analysis of RBI communications. First, as a single-equation method, LPs are more convenient when specifying the full system is challenging due to data limitations or model complexity. Second, LPs provide a more flexible framework for incorporating non-linearities or state-dependence, which may be crucial when analyzing the nuanced effects of central bank communications across different economic topics. Third, LPs are well-suited for estimating cumulative responses and multipliers, which are key to understanding the full impact of monetary policy statements. Given these advantages, and the fact that LPs and VARs are asymptotically equivalent under mild conditions, we opt for the LP approach in our analysis of RBI communications and their impact on Indian financial markets.

In conclusion, our work is trying to bridge some gaps in the existing literature. First, we provide the most extensive analysis to date of RBI communications, covering a crucial decade of evolution in India's monetary policy framework. Second, we apply state-of-the-art NLP techniques like BERTopic and sentiment analysis models fine-tuned on central bank communications to comprehensively analyze RBI communications and their market impacts. Third, we employ local projection methods (\cite{jorda2005}; \cite{jorda2024}) to estimate the dynamic effects of RBI communications on Indian financial markets. In doing so, we contribute to the broader literature on central bank communications while providing specific insights into the dynamics of monetary policy and financial markets in India.

\section{Data Collection and Preparation}
\label{sec:data}

\textbf{Sources of Data.} This section outlines the approach taken to collect, extract, and prepare data from the Reserve Bank of India's (RBI) Monetary Policy Committee (MPC) communications. The process is designed to systematically gather text data from two primary sources:

\begin{enumerate}
    \item Monetary Policy Statements (MPS)
    \item Transcripts of Q\&A sessions from Press Conferences given by the Governor and Deputy Governors of RBI.
\end{enumerate}

These documents are publicly available on the RBI's website, covering the period from April 2014 to June 2024. This comprehensive dataset provides a decade-long view of the RBI's Monetary Policy Communications.

\textbf{Data Extraction Process.} The process begins with the retrieval of documents from the RBI's official website. A series of specialized Python scripts were developed to automatically fetch these documents (see Appendix A for details). These scripts are designed to efficiently and accurately retrieve all relevant communications, including Monetary Policy Statements and Press Conference Transcripts, each identified by its unique URL or file location on the RBI website.

\textbf{Text Pre-Processing.}
Once the documents are retrieved, the text undergoes a series of pre-processing steps, also implemented through custom Python scripts. For web-based content, the scripts parse the HTML to extract relevant paragraphs, carefully excluding headers and other non-content elements. For PDF documents, a dedicated script employing PDF text extraction libraries is used to preserve the document's structure and ensure proper text ordering, especially for multi-column layouts.

The extracted text is then cleaned using additional scripted processes, removing superfluous elements like footnote markers, consolidating whitespace, and standardizing line breaks. This scripted cleaning process ensures that the text from all document types is transformed into a uniform format, facilitating subsequent analysis.

\textbf{Text Structuring.}
The cleaned text is structured according to its document type. For Monetary Policy Statements, the text is divided into individual paragraphs. For Q\&A transcripts, the text is separated into distinct question-answer pairs, with each speaker's contribution clearly delineated. Finally, the answers of the Governor and Deputy Governors, from the Q\&A exchange, are selected for the purpose of the analysis. 

\textbf{Metadata Association.}
Each textual unit (paragraph or answer) is assigned a unique identifier and associated with metadata including: Document type (MPS or Q\&A transcript), Publication date and Speaker information (for Q\&A transcripts). This structuring allows for multi-dimensional analysis across time, document types, and speakers. The final stage involves consolidating the extracted and pre-processed data into a unified dataset.

\section{\large Methodology}
\label{sec:methodology}

This section outlines the approach used to identify and analyze different economic topics in the Reserve Bank of India's Monetary Policy Committee (MPC) communications. Our methodology employs advanced natural language processing (NLP) techniques, primarily focusing on topic modeling, to extract meaningful insights from the textual data. Topic modeling is particularly relevant for analyzing RBI's MPC communications as it systematically uncovers latent themes, enabling the quantitative analysis of evolving policy priorities and concerns.

To implement this sophisticated analysis, we leverage the power of neural networks, a class of machine learning models inspired by biological neural systems. These networks comprise layers of interconnected nodes capable of learning intricate patterns from data. In particular, we rely heavily on Transformer models, a specialized neural network architecture designed to process sequential data such as text. These models employ an `attention' mechanism to focus on relevant input elements, enhancing their ability to comprehend contextual relationships in language. This advanced approach allows us to capture nuanced economic themes and policy sentiments that might be missed by traditional text analysis methods.

\subsection*{\large Topic Modeling with BERTopic}

Building on the foundation of neural networks and Transformer models discussed above, we employ a specific technique known as topic modeling to analyze the RBI's MPC communications. Topic modeling is a statistical method for uncovering abstract themes within a corpus of documents. This technique helps organize and summarize large volumes of textual data by identifying recurring patterns or topics, thereby aiding the analysis of complex textual datasets. Our approach builds upon BERT (Bidirectional Encoder Representations from Transformers), a state-of-the-art natural language processing model. 

BERT's key innovation lies in its ability to understand contextual relationships in text by considering context both before and after each word in a sentence. This architecture enhances the model's capacity to focus on relevant textual elements, improving its understanding of linguistic context. Before analysis, the text is tokenized, segmenting the input into smaller units—typically words or subwords—that serve as the foundational elements for the model's analysis.

To leverage BERT's capabilities for our specific task of identifying key economic topics in the RBI's communications, we employ BERTopic, an advanced topic modeling technique. BERTopic extends BERT's strengths by integrating transformer-based language models with class-based Term Frequency-Inverse Document Frequency (c-TF-IDF) to identify and cluster similar topics within text data (see Appendix B for details).

The c-TF-IDF component of BERTopic is a statistical measure that evaluates the importance of a word within a document relative to a collection or corpus (see Appendix B.1). This approach is particularly well-suited to our research, as it captures subtle differences in economic discussions that might be overlooked by traditional topic modeling techniques. By combining BERT's contextual understanding with c-TF-IDF's focus on term importance, BERTopic provides a powerful tool for uncovering nuanced economic themes in central bank communications.

To apply this sophisticated methodology to our research question, we begin our analysis with pre-processed sentence-level data from Monetary Policy Statements (MPS) and responses by the Governor and Deputy Governors during Press Conferences. These data are captured in the transcripts with Media, and Researchers and Analysts, covering the period from April 2014 to June 2024. This comprehensive dataset forms the foundation for our in-depth analysis of RBI's communication patterns over time.

\subsubsection*{Embedding Generation}

The first step in our BERTopic implementation is the generation of embeddings. Embeddings are dense vector representations of words or sentences that capture semantic meaning in a high-dimensional space. In our analysis, we pre-calculate sentence embeddings using the SentenceTransformer model, following best practices in BERTopic implementation.

The SentenceTransformer model generates numerical representations for entire sentences, enabling the quantification of semantic similarity between them, which facilitates subsequent analysis (see Appendix B.2). We selected this model for its optimal balance of performance and efficiency, as it effectively captures semantic nuances in financial language while remaining computationally feasible for our large dataset. Pre-calculation of embeddings also allows for faster iteration in subsequent steps of the analysis.

\subsubsection*{Dimensionality Reduction and Clustering}

While the embeddings generated in the previous step provide rich semantic information, they typically exist in a high-dimensional space, often comprising hundreds of dimensions. This high dimensionality reflects the model's attempt to capture the complex semantic relationships inherent in language. However, it also poses challenges for direct visualization and processing.

To address this complexity and make our data more manageable for analysis, we apply two key techniques: UMAP (Uniform Manifold Approximation and Projection) for dimensionality reduction, followed by HDBSCAN (Hierarchical Density-Based Spatial Clustering of Applications with Noise) for clustering.

We chose UMAP over alternatives like t-SNE or PCA because it better preserves both the local and global structure of the high-dimensional space (see Appendix B.3). This preservation is crucial for identifying coherent economic topics that may have complex inter-relationships. In the context of economic topics, preserving both local and global structure is particularly important because:

\begin{enumerate}
    \item Local structure preservation allows for the identification of nuanced differences between closely related economic concepts (e.g., different types of inflation measures).
    \item Global structure preservation maintains the overall relationships between broader economic themes (e.g., the connection between monetary policy, inflation, and economic growth).
    \item The combination of local and global preservation helps in capturing the hierarchical nature of economic topics, where specific subtopics nest within broader economic concepts.
\end{enumerate}

Following the dimensionality reduction, we employ HDBSCAN for clustering. This algorithm was selected for its ability to identify clusters of varying densities and shapes, a feature particularly useful in topic modeling where topic distributions may be heterogeneous. By using HDBSCAN, we can effectively group similar documents together, forming coherent topics that reflect the diverse range of economic themes present in RBI communications.

\subsubsection*{Topic Representation}

While clustering helps us identify distinct topics, interpreting these topics requires additional techniques to extract meaningful information. To effectively interpret the topics identified through our analysis, we utilize representation models, which are methods designed to highlight the key elements that define each topic. These models allow us to distill complex textual data into more understandable and meaningful representations. In this study, we employ multiple techniques to enhance the interpretability of the identified topics.

One such technique is \textbf{Maximal Marginal Relevance}. This approach ensures diversity in the selected phrases, reducing redundancy and capturing the multifaceted nature of economic discussions. By prioritizing both relevance and diversity, we can provide a more comprehensive representation of each topic.

Another key technique we employ is \textbf{Part-of-Speech Analysis}. By focusing on specific grammatical categories, this approach highlights relevant economic terminology. For example, it may emphasize nouns like ``inflation" or ``growth," and adjectives like ``fiscal" or ``monetary." This targeted analysis helps us zero in on the most pertinent economic concepts within each topic.

This multi-faceted approach provides a more comprehensive understanding of each identified topic, which is crucial for accurately mapping the diverse economic themes in RBI communications. Ensuring diversity in these phrases allows us to capture the breadth of ideas within a topic, avoiding over-simplification. The part-of-speech analysis helps highlight the most relevant economic terms and concepts, providing insight into the specific language used in policy discussions.

Combining these varied techniques allows us to not only capture the broad policy stances but also delve into specific sectoral concerns. This nuanced understanding enhances our ability to interpret subtle shifts in policy focus and communication themes over time. This is a critical aspect in the analysis of central bank communications, where even slight variations in language can have significant implications for economic agents and market participants.

\subsubsection*{Topic Model Training}

To explore the economic themes present in the data, we train the BERTopic model using the pre-calculated embeddings (see Appendix B.4 for details). Our goal is to generate approximately 10 distinct topics. This number was carefully chosen to strike a balance between granularity and interpretability, allowing us to capture the nuanced economic themes while keeping the analysis manageable.

 This number allows us to capture key areas such as inflation, growth, monetary policy, financial stability, and external sector developments, while also leaving room for more nuanced or emerging themes. This approach enables us to balance between capturing the complexity of monetary policy discussions and maintaining interpretability for our analysis. Furthermore, this number of topics provides sufficient granularity to track potential shifts in policy focus over time, which is crucial for addressing our research questions about how sentiment varies across different economic topics and what this variation reveals about India's monetary policy priorities.

\subsection*{\large Sentiment Analysis}

Sentiment analysis is a computational technique used to determine the emotional tone behind a series of words. In the context of central bank communication, such as monetary policy statements, and press releases, sentiment analysis helps quantify the tone or sentiment conveyed by the central bank. This analysis can provide insights into the attitudes, confidence levels, and concerns expressed by central bankers regarding economic conditions, policy decisions, and future outlooks.

In monetary policy, central banks like the Reserve Bank of India (RBI) use specific language that can signal their policy intentions to the market. These signals are crucial because they influence market expectations, investor behavior, and economic outcomes. Sentiment analysis, therefore, becomes a powerful tool for interpreting these signals. It categorizes the language used in these communications into three primary sentiments: `Dovish,' `Hawkish,' and `Neutral' (see Appendix C for examples).

\textbf{Dovish Sentiment.} It reflects an accommodative or expansionary monetary policy stance. This language indicates that the central bank is likely to lower interest rates or implement measures to stimulate economic growth. Dovish statements often emphasize the need for supportive monetary policies, especially during periods of low inflation, high unemployment, or economic downturns. Markets may react to dovish communications by expecting continued or increased monetary stimulus, which can result in lower bond yields, a weaker currency, and rising stock prices.

\textbf{Hawkish Sentiment.} It indicates a tighter or more restrictive monetary policy stance. This suggests that the central bank might raise interest rates or reduce liquidity to control inflation or prevent the economy from overheating. Hawkish language typically focuses on the risks of high inflation or financial instability, signaling that the central bank prioritizes price stability even at the potential cost of slower economic growth. Markets often interpret hawkish statements as a signal of upcoming rate hikes or tighter monetary conditions, leading to higher bond yields, a stronger currency, and potentially lower stock prices

\textbf{Neutral Sentiment.} It reflects a balanced or unchanged policy stance. In central bank communications, this might involve stating current economic conditions and policy measures without indicating any immediate changes. Neutral statements often focus on monitoring economic indicators, explaining how monetary policy is being transmitted through the economy without indicating a clear directional bias, or reiterating the status quo. Neutral sentiment is important because it signals to markets that the central bank is confident in the current policy direction and does not foresee significant changes in the near term.

We utilize a state-of-the-art language model, RoBERTa (Robustly Optimized BERT Pre-training Approach), to perform this task. RoBERTa, developed by \cite{liu2019} \footnote{RoBERTa, an optimized version of BERT, utilizes a transformer architecture with 24 layers, 16 attention heads, and 355M parameters, trained on a 160GB text corpus (Liu et al., 2019).}, is an optimized version of BERT (Bidirectional Encoder Representations from Transformers) that has shown superior performance in various natural language processing tasks, including sentiment analysis.

\textbf{\large RoBERTa Model.} RoBERTa is a transformer-based model that learns contextual relations between words in a text. It is pre-trained on a large corpus of text and can be fine-tuned for specific tasks such as sentiment classification. In our study, we use a RoBERTa model that has been specifically fine-tuned for analyzing financial texts, which makes it particularly suitable for interpreting central bank communications.

Our sentiment analysis process involves the following steps:

\textbf{Text Tokenization.} The first step is text tokenization, where the input text is broken down into smaller units (tokens), such as words or subwords, that can be processed by the model. The RoBERTa tokenizer, which is used in this process, converts the text into a numerical format that preserves the meaning and context of each token within the sentence.

\textbf{Model Prediction.} After tokenization, the processed text is passed through the pre-trained RoBERTa model. In this case, the model generates probability scores for each sentiment category—Dovish, Hawkish, and Neutral—based on the patterns and contextual relationships it identifies in the text.

\textbf{Sentiment Assignment.} Finally, the sentiment with the highest probability score is assigned to the text. For example, if the model determines that the text has a higher probability of being classified as Hawkish compared to Dovish or Neutral, it will be labeled as Hawkish. This classification allows researchers to quantify and track the sentiment of central bank communications over time.

\textbf{Sentiment Measure Calculation.}  
In addition to the sentiment assignment, we employ a more nuanced approach to capture both the direction and intensity of sentiment. For each group of communications (grouped by date and topic), we calculate:

\begin{itemize}
    \item \textbf{Average Sentiment Score:} The mean sentiment score for all sentences within a group, offering an aggregate measure of tone.
    \item \textbf{Sentiment Balance:} A ratio representing the relative prevalence of dovish versus hawkish sentiment, calculated as (\cite{gorodnichenko2023}):
    \[
    \text{Sentiment Balance} = \frac{\text{Dovish Count} - \text{Hawkish Count}}{\text{Dovish Count} + \text{Hawkish Count}}
    \]
    This metric captures shifts in the overall sentiment balance within a given topic.
\end{itemize}

These measures are calculated at both the sentence level and aggregated across entire policy statements or press conference transcripts, providing a comprehensive view of the sentiment dynamics in RBI communications. Sentences are further categorized into predefined topics, such as ``Banking Sector and Payment Systems," ``Inflation Dynamics and Monetary Policy Framework," to calculate topic-specific sentiment scores. These multi-level sentiment scores serve as key independent variables in our subsequent analysis of financial market reactions.

In summary, our methodology combines advanced NLP techniques to provide a comprehensive examination of RBI's monetary policy communications. We begin by employing BERTopic, a state-of-the-art topic modeling technique, to identify key economic themes in MPC statements and press conference transcripts. This process involves generating sentence embeddings, reducing dimensionality while preserving important structural relationships, and clustering similar content. We then use multiple representation models to interpret these topics effectively. By applying this methodology to our dataset spanning from 2014 to mid-2024, we can track the evolution of communication themes over time, offering insights into how the RBI's policy priorities and communication strategies have adapted to India's changing economic landscape. Following this, we conduct a detailed sentiment analysis to quantify the tone of these communications. The results, as presented in the following section, detail the identified topics and their prevalence over time, before discussing their implications for understanding India's monetary policy communication and its impact on financial markets.

\section{Data Analysis and Visualization}
Our analysis of RBI's Monetary Policy Committee (MPC) communications from April 2014 to June 2024 reveals several key patterns in topic distribution and sentiment evolution. We employ various visualization techniques to elucidate these patterns.
\subsection{Topic Distribution and Evolution}
\begin{figure}
\centering
\includegraphics[width=0.99\linewidth]{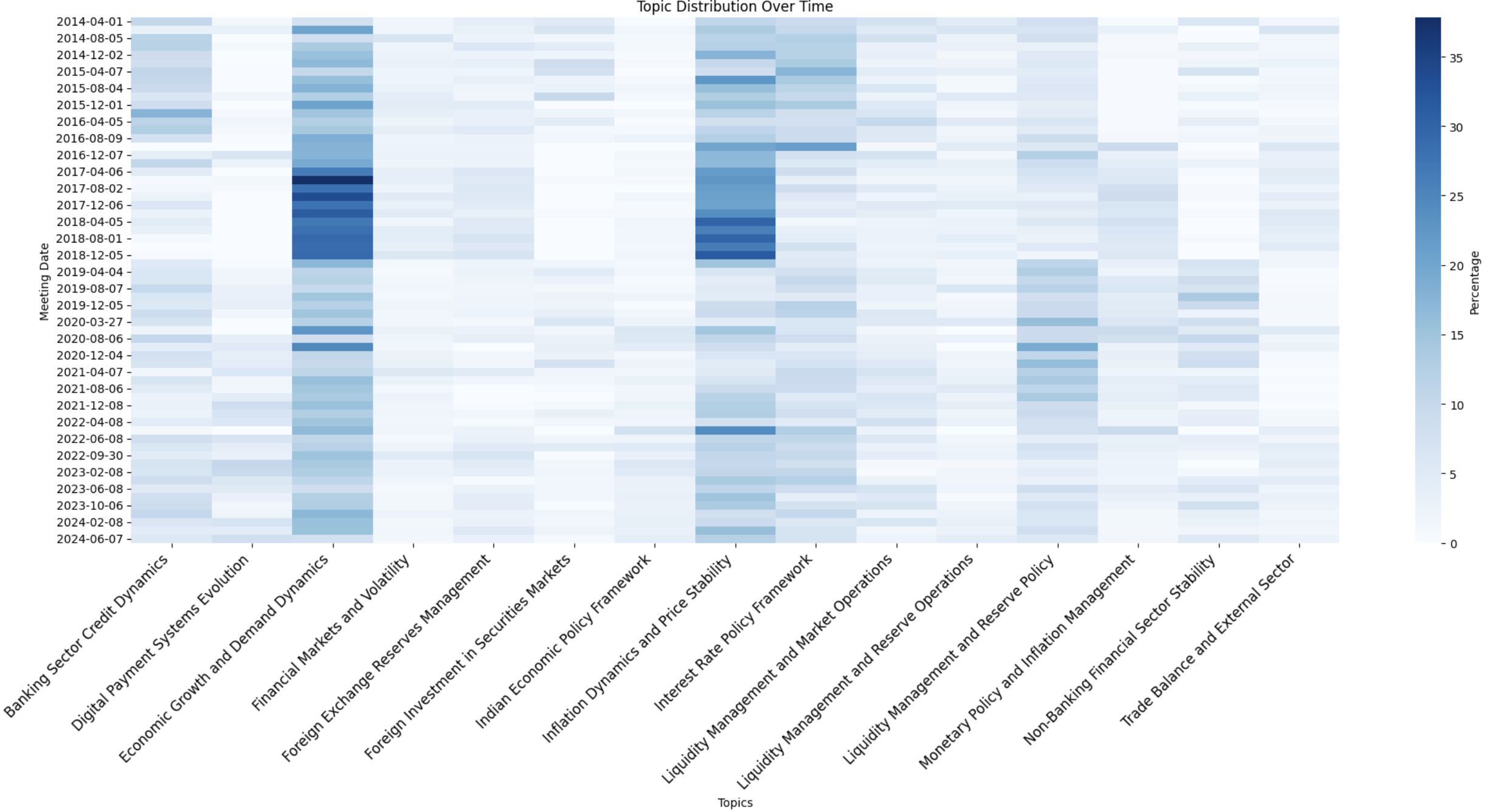}
\caption{Topic Distribution Heatmap Across MPC Meetings (2014-2024)}
\label{fig:topic_heatmap}
\end{figure}

Figure \ref{fig:topic_heatmap} presents a heatmap of topic distribution across MPC meetings. This visualization reveals the shifting focus of RBI communications over time, reflecting changing macroeconomic conditions and policy priorities.
The heatmap indicates that certain topics, such as ``Economic Growth and Demand Dynamics" and ``Inflation Dynamics and Price Stability," maintain a consistent presence throughout the period. However, we observe notable shifts in emphasis. For instance, ``Digital Payment Systems Evolution" gains prominence in later years, reflecting India's rapid digitalization of financial services.
Figure 3 provides a more granular view of topic evolution over time. This line graph allows us to track the relative importance of each topic across our study period.
\begin{figure}
\centering
\includegraphics[width=0.99\linewidth]{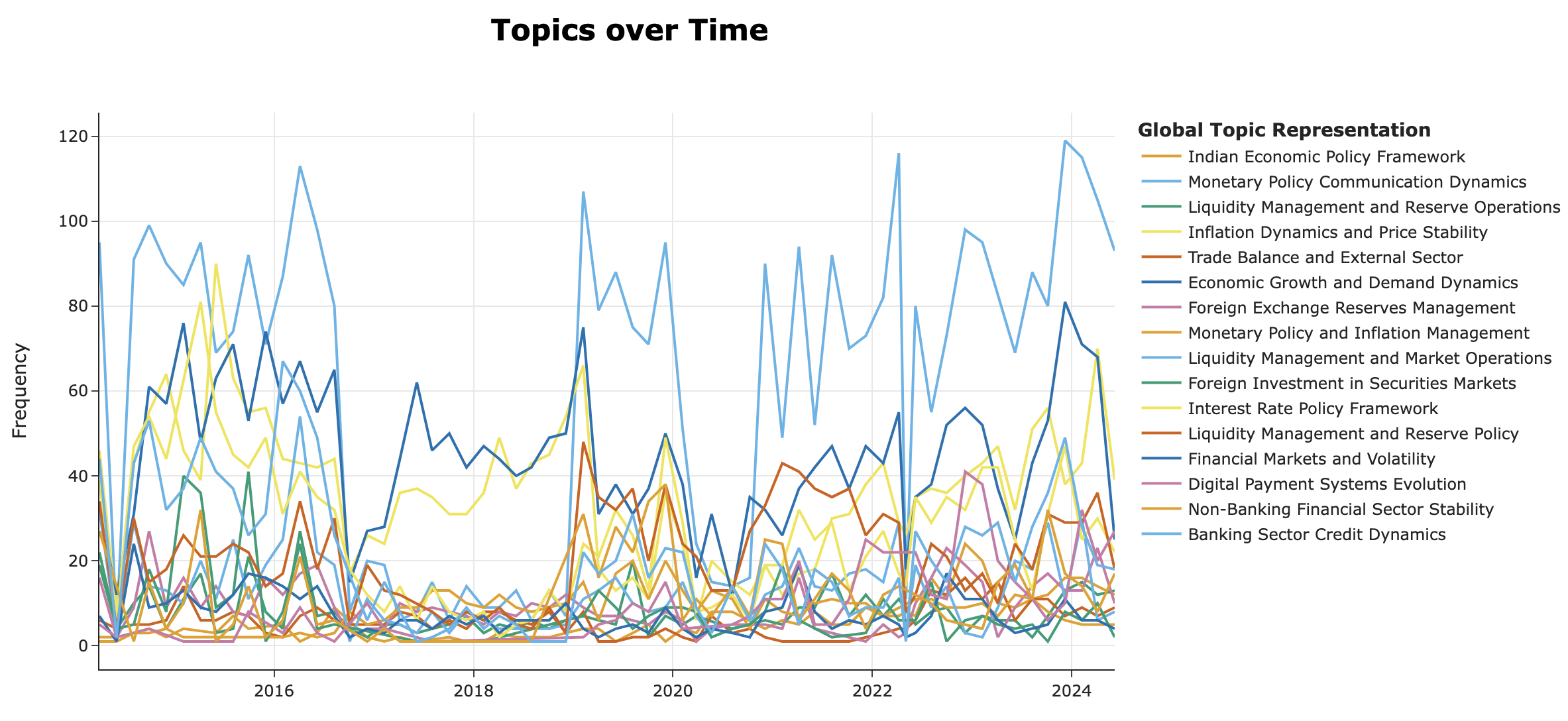}
\caption{Evolution of Topic Prevalence in RBI Communications (2014-2024)}
\label{fig:topics_over_time}
\end{figure}
The graph reveals cyclical patterns in some topics, such as ``Trade Balance and External Sector," which may correspond to India's evolving position in global trade. We also observe a marked increase in discussions related to ``Liquidity Management and Reserve Operations" from 2020 onwards, likely reflecting the RBI's response to the economic challenges posed by the COVID-19 pandemic.
\subsection{Document-Topic Relationships}
To understand how individual communications relate to our identified topics, we present a scatter plot of documents in reduced dimensional space (Figure \ref{fig:hawkish_sentiment}).
\begin{figure}
\centering
\includegraphics[width=1\linewidth]{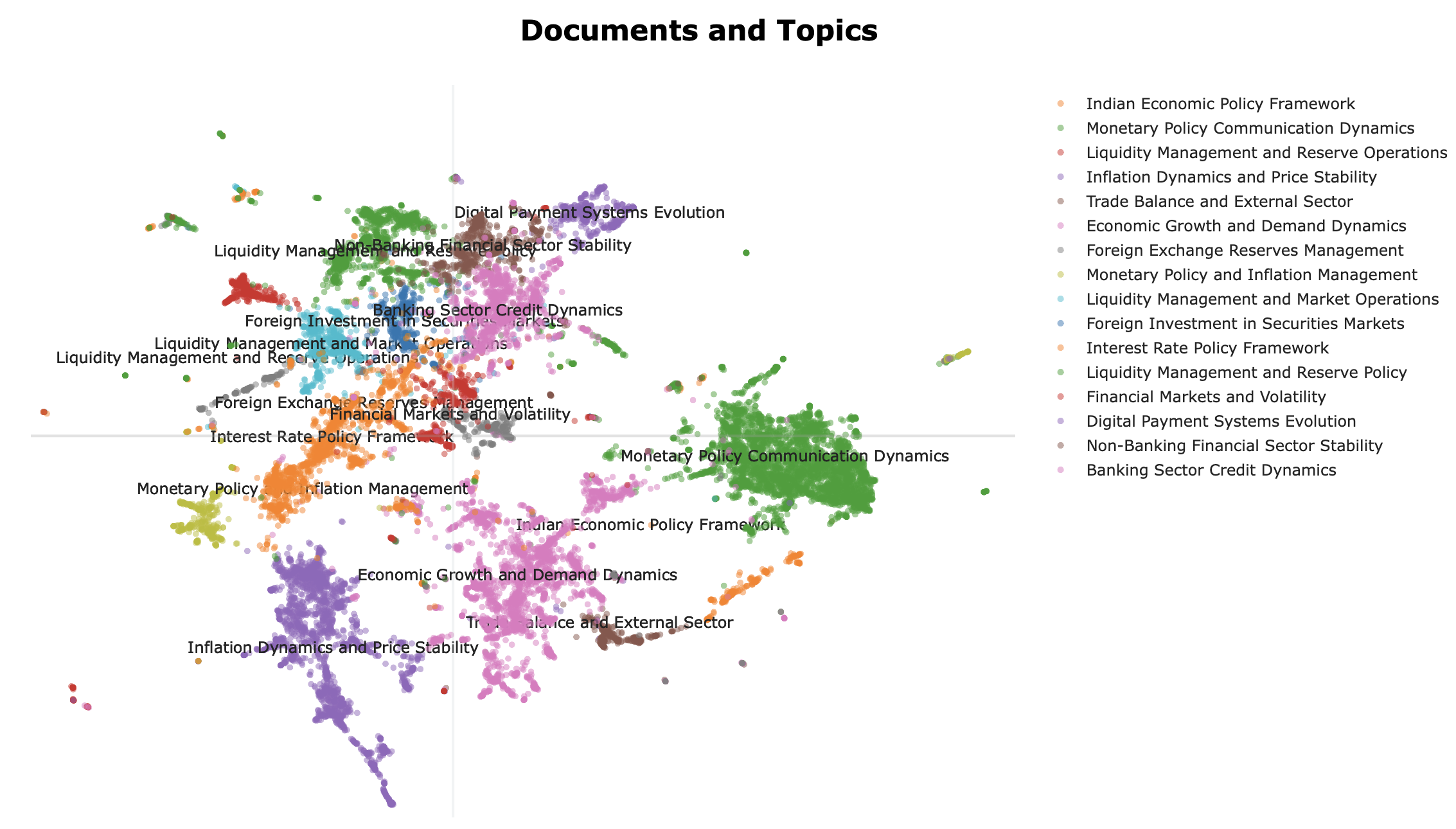}
\caption{Document-Topic Distribution in Reduced Dimensional Space}
\label{fig:doc_topic_map}
\end{figure}
This visualization reveals distinct clusters of documents, with some topics forming tight groups while others show more dispersion. The proximity of clusters related to ``Inflation Dynamics" and ``Monetary Policy" suggests a close relationship between these themes in RBI communications, consistent with the central bank's inflation-targeting mandate.
\subsection{Sentiment Analysis}
\begin{figure}
\centering
\includegraphics[width=0.9\linewidth]{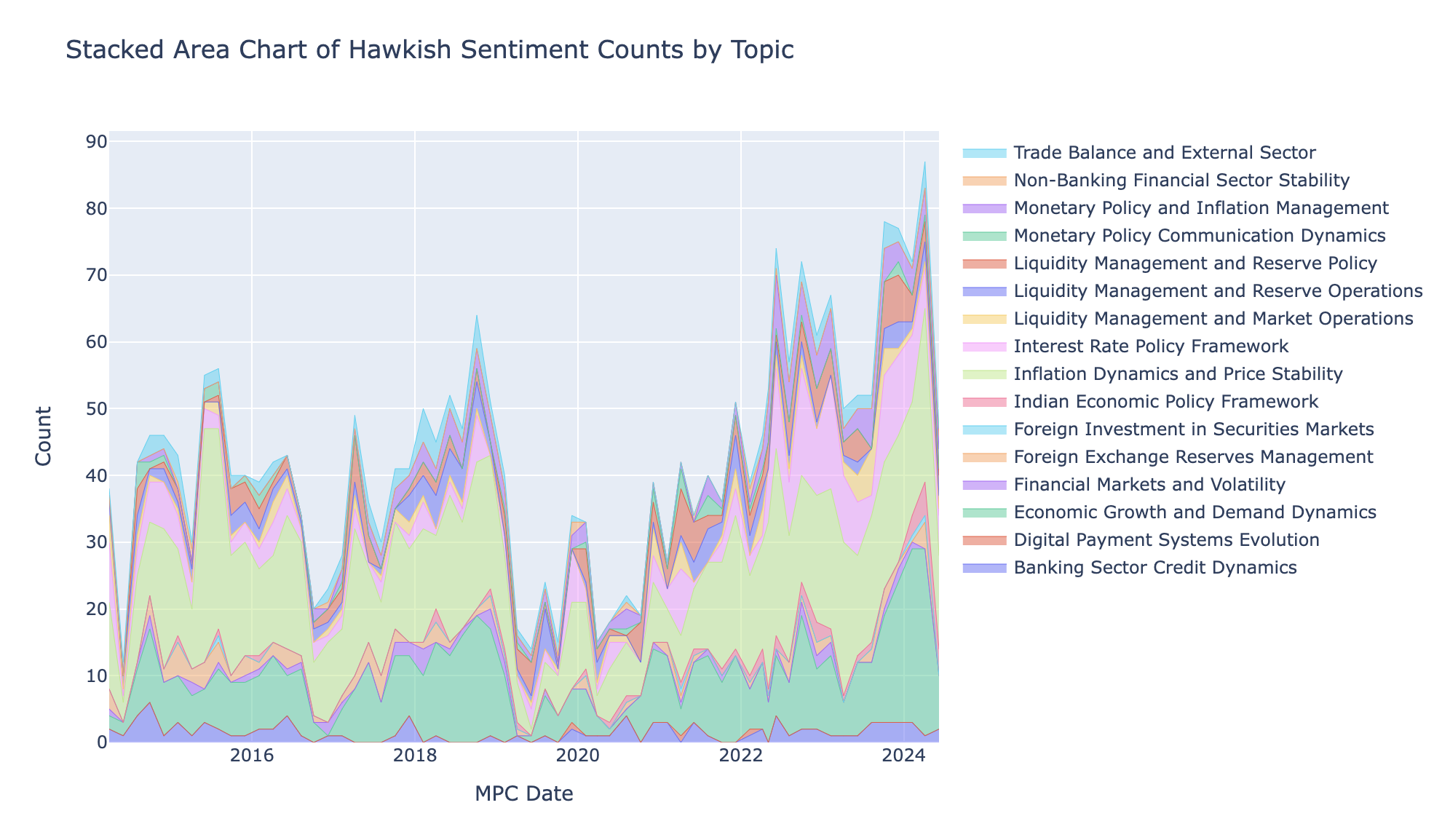}
\caption{Distribution of Hawkish Sentiment Across Topics (2014-2024)}
\label{fig:hawkish_sentiment}
\end{figure}

\begin{figure}
\centering
\includegraphics[width=0.9\linewidth]{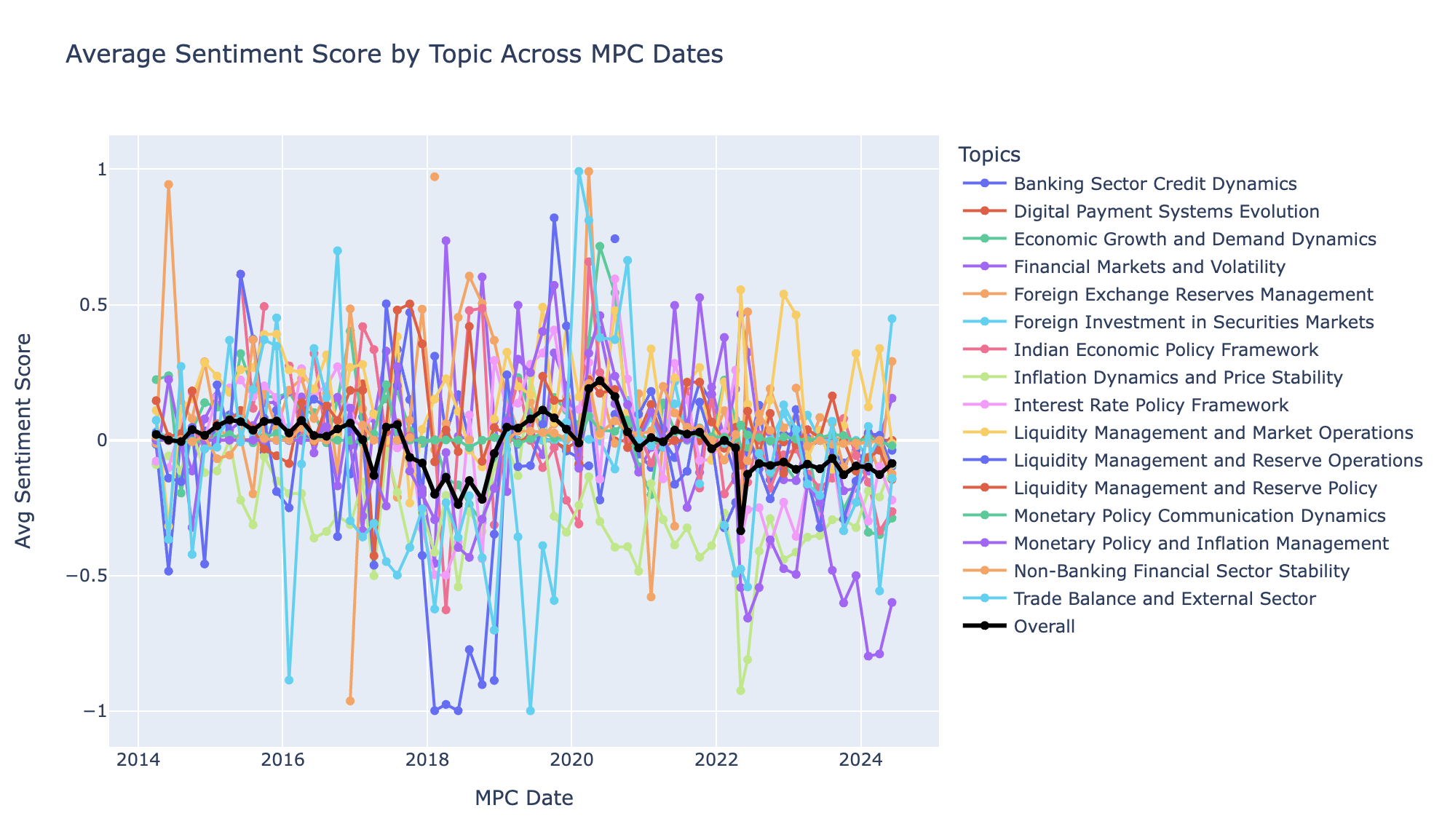}
\caption{Average Sentiment Scores by Topic in RBI Communications}
\label{fig:avg_sentiment}
\end{figure}

Our sentiment analysis provides insights into the tone of RBI communications across different topics. Figure \ref{fig:hawkish_sentiment} illustrates the distribution of hawkish sentiment across topics over time.
We observe that hawkish sentiment is particularly pronounced in discussions of ``Inflation Dynamics and Price Stability," especially during periods of elevated inflation. Conversely, topics related to ``Economic Growth and Demand Dynamics" tend to exhibit less hawkish sentiment, potentially reflecting the RBI's dual mandate of price stability and economic growth.
Figure \ref{fig:avg_sentiment} presents the average sentiment scores across topics, providing a broader view of the RBI's communication tone.
This visualization reveals that some topics, such as ``Digital Payment Systems Evolution," consistently receive more positive sentiment, possibly reflecting the RBI's optimistic outlook on financial innovation. In contrast, topics related to external sector dynamics show more variation in sentiment, likely mirroring the volatility of global economic conditions.

\begin{figure}
\centering
\includegraphics[width=0.9\linewidth]{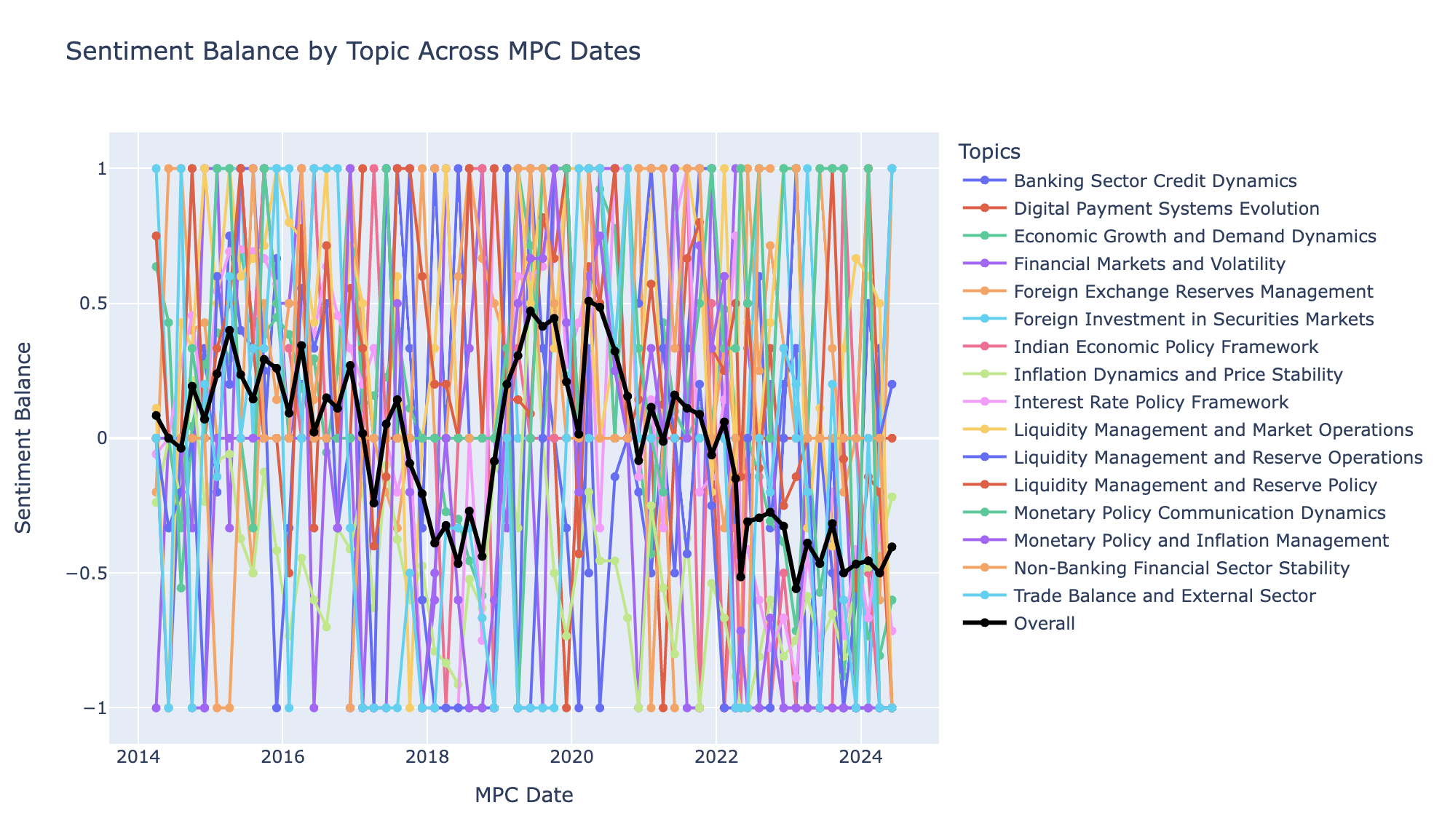}
\caption{Sentiment Balance by Topic in RBI Communications}
\label{fig:sentiment_balance}
\end{figure}

\begin{figure}
\centering
\includegraphics[width=0.99\linewidth]{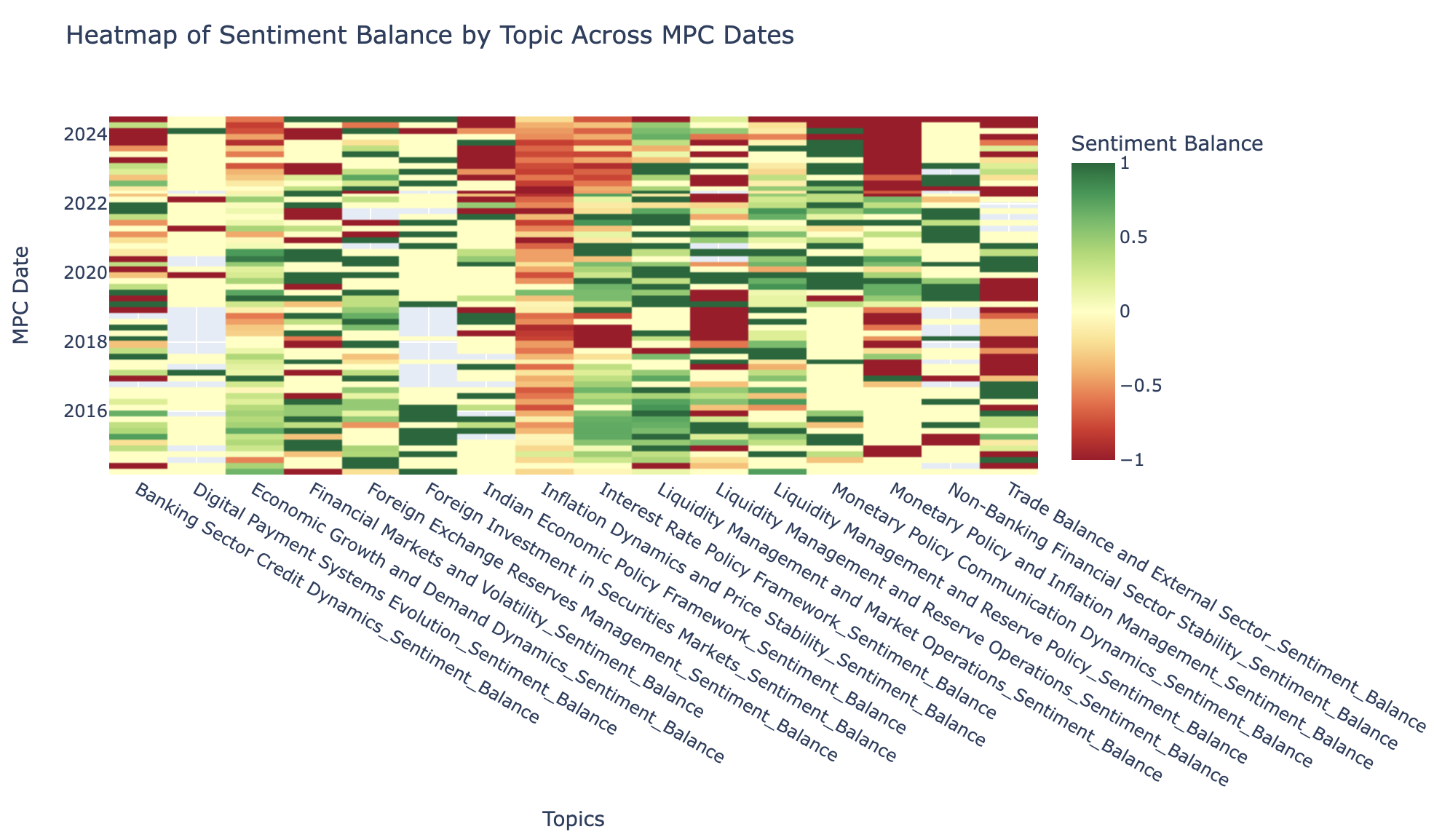}
\caption{Sentiment Heatmap Across Topics and Time in RBI Communications}
\label{fig:sentiment_heatmap}
\end{figure}

To capture the nuanced balance between dovish and hawkish sentiments, we calculate a sentiment balance score for each topic (Figure \ref{fig:sentiment_balance}).
This metric reveals that while some topics maintain a relatively stable sentiment balance over time, others exhibit significant fluctuations. For instance, discussions of ``Monetary Policy and Inflation Management" show a shift towards more hawkish sentiment in recent years, possibly indicating a tightening bias in response to inflationary pressures.
Finally, Figure \ref{fig:sentiment_heatmap} presents a heatmap of sentiment across topics and time, offering a comprehensive view of how the tone of RBI communications has evolved.
This visualization allows us to identify periods of policy shift, such as the notably dovish tone across multiple topics in early 2020, likely reflecting the RBI's accommodative stance in response to the COVID-19 pandemic.
In summary, our analysis reveals significant variation in both the content and sentiment of RBI communications across topics and time. These patterns provide valuable insights into the evolving priorities of Indian monetary policy and the RBI's response to changing economic conditions.

\section{Empirical Strategy and Findings}

\subsection*{Empirical Strategy}

To analyze the impact of RBI communications on Indian financial markets, we employ the Local Projections method as developed by  \cite{jorda2005}. This approach allows us to estimate the dynamic effects of sentiment shocks in RBI's communications on financial market variables over different time horizons. Our baseline specification is as follows:

\begin{equation}\label{eq:local_projection}
y_{i,t+h} - y_{i,t}= \beta_0^{(h)} + \beta_1^{(h)} S_{j,t}  + \beta_2^{(h)} X_t + \beta_3^{(h)} \text{COVID}_t + + \beta_4^{(h)} \text{Rajan}_t + \beta_5^{(h)} \text{Patel}_t  +\varepsilon_{i,t+h}
\end{equation}

where $t$ denotes RBI MPC meeting dates. $y_{i,t+h}$ is the log price of financial asset $i$ at horizon $h$, $y_{i,t}$ is the log price of financial asset $i$ on the day of the RBI MPC meeting, $S_{j,t}$ is the sentiment score for cluster $j$ at time $t$, and $X_t$ is a vector of control variables. $\text{COVID}_t$ is a dummy variable equal to 1 for dates after March 11, 2020, when WHO declared COVID-19 a pandemic, and 0 otherwise (after December 31, 2021). $Rajan_t$
and $Patel_t$ are dummy variables corresponding to the tenures of Governors Raghuram Rajan and Urjit Patel, respectively. The tenure of Governor Shaktikanta Das, the longest-serving governor during the period under study, is taken as the reference category. $h$ is the horizon (0 to 30 days), and $\varepsilon_{i,t+h}$ is the error term.

We estimate the above specification separately for each horizon $h$ and plot the estimated coefficients, e.g., $\left\{ \hat{\beta}_1^{(h)} \right\}_{h=0}^{30}$ to illustrate the dynamics of the response to (text) sentiment in RBI communications. Our analysis primarily focuses on daily data to capture the longer-term effects of sentiment on financial market, which might be critical for understanding sustained impacts.

The outcome variable in our study is the daily financial indicator, represented by the (closing/opening) price of the BSE Sensex, sourced from the National Stock Exchange (NSE) and other financial data providers such as Yahoo Finance and FRED. The BSE Sensex, widely regarded as a benchmark index for the Indian equity market, is employed to gauge the stock market's reaction to monetary policy communications. The return is calculated as the log difference between the closing and opening prices, i.e.,

$y_{i,t+h} - y_{i,t} = \log(BSESN^{\text{close}}_{t+h}) - \log(BSESN^{\text{open}}_{t}),$

where $t$ denotes the day of the MPC meeting. Hence, on the day of the MPC meeting, the return is captured by the log difference between the BSE Sensex’s close and open prices.

Given the limited sample size (approximately 64 MPC meetings within the study period), we employ bootstrap methods to address potential biases in our estimates and to construct confidence intervals with high coverage. In order to do this, we generate 90 percent confidence intervals based on the 5th and 95th percentiles of the estimates derived from bootstrap samples. 

We estimate equation \ref{eq:local_projection} separately for each sentiment cluster $j$ to isolate the impact of sentiment in different economic topics. The clusters we analyze are: Inflation Dynamics and Price Stability, Trade Balance and External Sector, Economic Growth and Demand Dynamics, Foreign Exchange Reserves Management, Foreign Investment in Securities Markets, Interest Rate Policy Framework, Financial Markets and Volatility and Banking Sector Credit Dynamics. 

Our control variable, $X_t$, includes `Monetary Policy Shock' following the methodology of \cite{lakdawala2024measuring}. \footnote{As shared on the following website \url{https://sites.google.com/view/rasesite/data}} \footnote{Due to the need to truncate RBI Communications data to align with MP shock series, these results have been shared in the Appendix E.}
In order to test the sensitivity of estimates, we have conducted robustness checks by using additional control variables such as Inflation Growth Rate and Policy Uncertainty Index.\footnote{Data for the inflation and policy uncertainty variable is sourced from the FRED Database.} We use heteroskedasticity and autocorrelation consistent (HAC) standard errors for inference. This approach accounts for potential serial correlation in the error terms, which is particularly important given the overlapping nature of our dependent variable at horizons greater than one day.

The coefficient of interest, $\beta^{h}$, represents the cumulative effect of a one-unit increase in the sentiment score for cluster $j$ on the financial asset price over horizon $h$. By estimating this effect for multiple horizons, we can trace out the dynamic response of financial markets to RBI communications across different economic topics.

The inclusion of the $\text{COVID}_t$ dummy variable allows us to control for the structural break in financial market behavior during the pandemic period, helping isolate the impact of RBI communications from the broader economic disruptions caused by COVID-19. Additionally, we include dummy variables for the tenures of Governors Raghuram Rajan and Urjit Patel, with Governor Shaktikanta Das serving as the reference category. This enables us to account for potential differences in market reactions attributable to changes in leadership at the RBI during the study period.

This empirical strategy enables us to address our key research questions:

\begin{enumerate}
\item How does sentiment in RBI's MPC communications vary across different economic topics?
\item How do these topic-specific sentiments impact Indian financial market?
\end{enumerate}

By examining the $\beta^{h}$ coefficients across different clusters and financial assets, we can gain insights into which aspects of RBI communications have the most significant market impact and how these effects evolve over time. 

Also, another thing to note is that we are conducting a `Sentiment-Wise Regression' analysis, that is, setting up individual regressions for each sentiment cluster instead of including all sentiments in a single model. This approach mitigates multi-collinearity, ensuring that the effects of each sentiment are estimated independently, without interference from related sentiments. It also enhances the clarity and interpretability of our findings, allowing us to isolate the impact of specific monetary policy topics on financial markets. Furthermore, this method reduces the risk of over-fitting and provides flexibility in model specification, enabling a more tailored and robust analysis. By focusing on individual sentiment clusters, we capture the nuanced effects of central bank communication, offering deeper insights into the varied reactions of financial markets.

 \subsection*{Findings}

\begin{figure}
    \centering
    \includegraphics[width=1\linewidth]{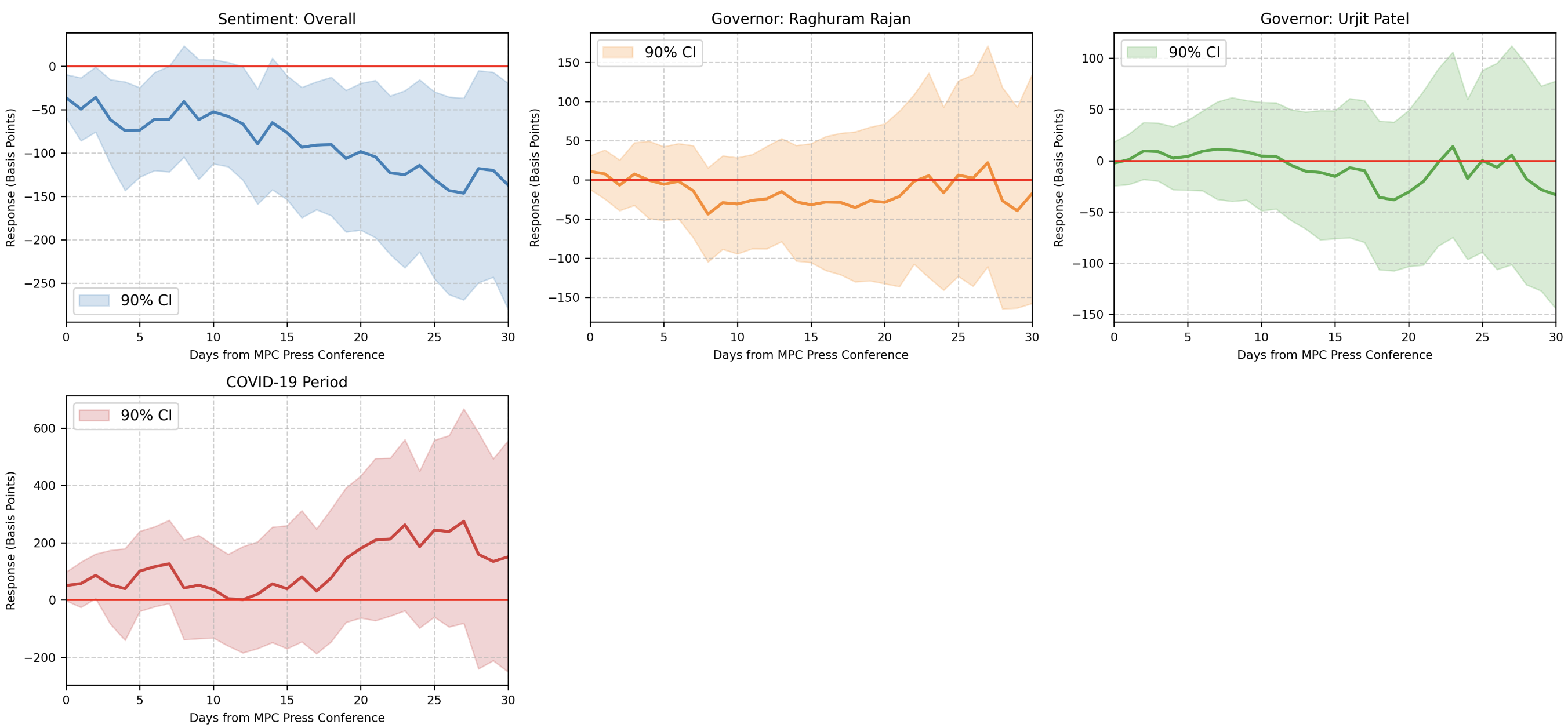}
    \caption{\textbf{Dynamic Responses of BSE Sensex (BSESN) to Aggregate (text) Sentiment} \newline \footnotesize This figure presents the dynamic responses of the BSE Sensex to aggregate sentiment from local projection regressions over a 30-day horizon following RBI Monetary Policy Committee meetings. The first panel shows the overall sentiment, while the remaining panels illustrate the responses specific to the tenure of Governors Raghuram Rajan and Urjit Patel (with the tenure of Shri Shaktikanta Das as reference), and during the COVID-19 period. The solid lines represent point estimates, with the shaded areas indicating 90\% bias-corrected bootstrap confidence intervals. These responses are derived from the baseline model, which includes governor dummies and the COVID-19 dummy to control for structural changes in financial markets. This allows us to disentangle the sentiment effect from potential confounding factors such as leadership changes at the RBI or pandemic-related disruptions.}
    \label{fig:bse-response}
\end{figure}

Our analysis of the Reserve Bank of India's (RBI) Monetary Policy Committee (MPC) communications reveals significant and persistent impacts on the Indian stock market, as measured by the BSE Sensex. We examine the effects using local projections to estimate cumulative responses over a 30-day horizon following MPC meetings \footnote{As a baseline, we estimate specification (1) with Sentiment Balance as the sole regressor to capture the direct relationship between sentiment and market returns without the potential confounding effects of additional variables. This approach allows us to observe the unadjusted impact of sentiment on financial markets included in Appendix E.} \footnote{In subsequent robustness checks, we explore the sensitivity of our estimates by incorporating additional control variables (e.g. Monetary Policy Shock) and considering alternative model specifications (e.g. Interaction Terms)}.

\subsubsection*{Impact of Text-based Aggregate Sentiment}

The aggregate sentiment of RBI (text) communications, measured by our sentiment score, shows a substantial and persistent negative effect on the BSE Sensex:
\begin{itemize}
\item A one-unit increase in dovish sentiment is associated with approximately a 150 basis point decrease in the Sensex after 30 days.
\item This effect builds gradually over time, becoming statistically significant after about 10 days and continuing to strengthen thereafter.
\item The negative response suggests that more dovish language in MPC communications tends to lead to a decline in the broader equity market.
\end{itemize}
This finding contrasts with the conventional expectation that dovish central bank communication typically boosts stock markets. Generally, lower interest rates, signaled by dovish monetary policies, are generally expected to boost stock markets by making borrowing cheaper and increasing the present value of future cash flows (\cite{bernanke2004conduct}).

It may indicate that market participants in India interpret dovish RBI communications as a signal of economic concerns rather than as positive news about easier monetary conditions.

\textbf{Magnitude and Persistence of the Effect} Several notable characteristics of the response warrant discussion. The effect is both large in magnitude and persistent in duration, with the cumulative response reaching approximately -150 basis points over the 30-day horizon. Unlike what might be expected for daily returns, we do not observe mean reversion in the cumulative returns (\cite{fama1965behavior}). This suggests that the initial impact of dovish sentiment is not quickly reversed by the market.

The continuous decline over the 30-day period goes against the common expectation that market reactions to news or policy changes stabilize or mean-revert over time. These observations align with the concept of slow-moving capital proposed by \cite{duffie2010asset} and further explored by \cite{fleckenstein2014treasury}. This theory suggests that not all investors can react immediately to new information, leading to gradual price adjustments rather than immediate jumps.

\textbf{Governor-Specific Effects} Our analysis reveals differences in market reactions across different RBI governors. During Dr. Raghuram Rajan's tenure, dovish communications had a negative but largely insignificant effect on the Sensex. Under Dr. Urjit Patel, the market response to dovish sentiment was initially positive but became negative after about 15 days, though the effect remains statistically insignificant throughout the 30-day horizon. The baseline effect (under current Governor Shaktikanta Das) is strongly negative, as described in the overall sentiment analysis above.

These differences suggest that market participants may interpret similar communication (text) tones differently depending on the governor in office.

\textbf{COVID-19 Period} The impact of RBI communications during the COVID-19 period shows a markedly different pattern. Dovish sentiment during this period is associated with a positive response in the Sensex, in contrast to the negative response seen overall. The effect becomes statistically significant after about 20 days and reaches approximately 250 basis points by day 30. This reversal in the direction of the effect during the COVID-19 period suggests that the unusual economic circumstances led market participants to interpret dovish RBI communications more conventionally – as a positive signal for equity markets.

\textbf{Implications} These findings have several important implications. The negative response to dovish sentiment in normal times suggests that RBI communications may be primarily interpreted as signals about the state of the economy rather than about future policy actions. This interpretation aligns with the ``information effect" discussed in central bank communication literature (\cite{nakamura2018high}).

The differences across governors highlight the importance of central banker credibility and consistency in communication. The reversal of the effect during COVID-19 underscores how extraordinary economic circumstances can fundamentally alter the interpretation of central bank communications. The gradual building of the effect over time, rather than an immediate jump, supports the slow-moving capital hypothesis and suggests that market participants take time to fully process and react to the nuances in RBI communications.

The persistent negative effect on the BSE index following dovish sentiment in RBI communications could have important implications for understanding how the Indian market interprets central bank communication. These results contribute to our understanding of how RBI communications impact Indian financial markets and highlight the complex, context-dependent nature of market responses to central bank communication. They underscore the need for careful, nuanced communication strategies that consider not just the immediate market reaction, but also longer-term effects and varying interpretations under different economic conditions.

\subsubsection*{Cluster-Specific Sentiment Effects}
Our analysis reveals heterogeneous impacts of sentiment across different economic topic clusters on the BSE Sensex. As before, we examine these effects using local projections to estimate cumulative responses over a 30-day horizon following MPC meetings.

\textbf{Banking Sector Dynamics}
\begin{figure}
\centering
\includegraphics[width=1\linewidth]{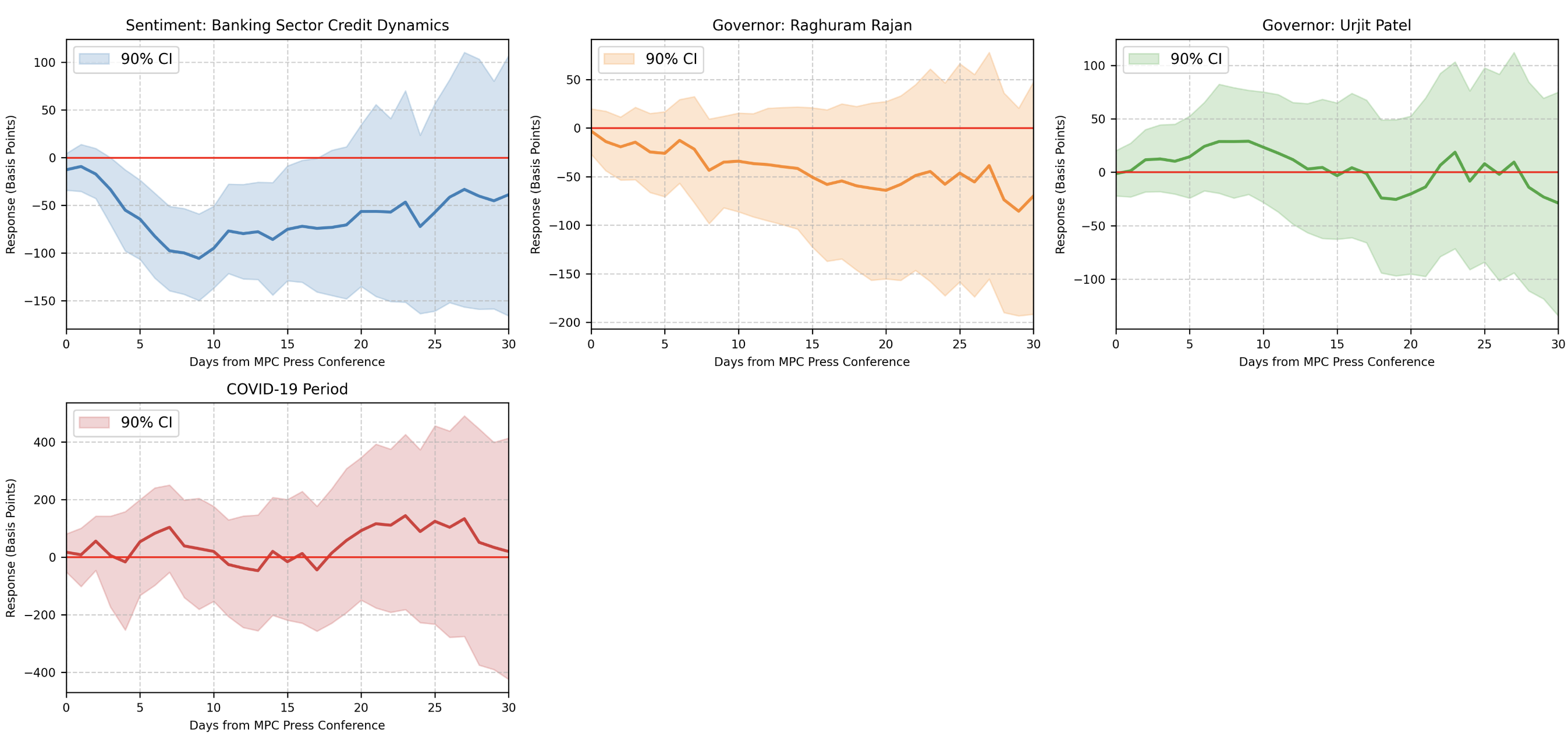}
\caption{Dynamic Responses of BSE Sensex (BSESN) to the cluster \textbf{`Banking Sector Dynamics'}}
\label{fig:banking-sector-dynamics}
\end{figure}
The impulse responses depicted in Figure \ref{fig:banking-sector-dynamics} illustrate the dynamic effects of sentiment related to Banking Sector Credit Dynamics on the BSE Sensex (BSESN) over a 30-day horizon following RBI Monetary Policy Committee (MPC) meetings. These results are derived from our local projections model, which accounts for COVID-19 pandemic and governor-specific effects.

Sentiment Effect:
The top-left panel shows the cumulative response of the BSE Sensex to a one-unit increase in the sentiment score for Banking Sector Credit Dynamics. We observe a consistently negative response over the 30-day horizon, with the effect becoming statistically significant after approximately 5 days. The magnitude of the effect peaks around day 10, with the Sensex decreasing by about 100 basis points. This suggests that more dovish sentiment regarding banking sector credit dynamics is associated with a decline in the stock market index.

The negative relationship between positive sentiment in banking sector communications and stock market performance may seem counter-intuitive at first. However, it could be interpreted in several ways. Market concerns about potential overheating in the banking sector when the RBI signals strong credit growth. Expectations of tighter monetary policy in response to robust credit expansion, which could negatively impact equity valuations.
Possible anticipation of increased regulatory scrutiny or macro-prudential measures in response to rapid credit growth.

Governor-Specific Effects:
The top-right and bottom-left panels show the differential effects during the tenures of Governors Raghuram Rajan and Urjit Patel, respectively, compared to the baseline period of Governor Shaktikanta Das.
During Dr. Rajan's tenure, we observe a generally negative differential effect, suggesting that the market response to banking sector sentiment was more pronounced compared to the Shri Das' period. This effect persists through the 30-day horizon.
For Dr. Patel's tenure, the differential effect is initially positive but becomes negative after approximately 15 days.

This pattern suggests a more complex market reaction during Dr. Patel's governorship, possibly reflecting evolving market perceptions or policy shifts during his tenure.
COVID-19 Impact:
The bottom-right panel captures the differential effect during the COVID-19 period. We observe a highly volatile response indicating substantial uncertainty in market reactions during this unprecedented time. The point estimates suggest a generally positive differential effect, implying that the negative impact of banking sector sentiment on the Sensex was somewhat mitigated during the pandemic. 

\textbf{Economic Growth and Demand Dynamics}
\begin{figure}
\centering
\includegraphics[width=1\linewidth]{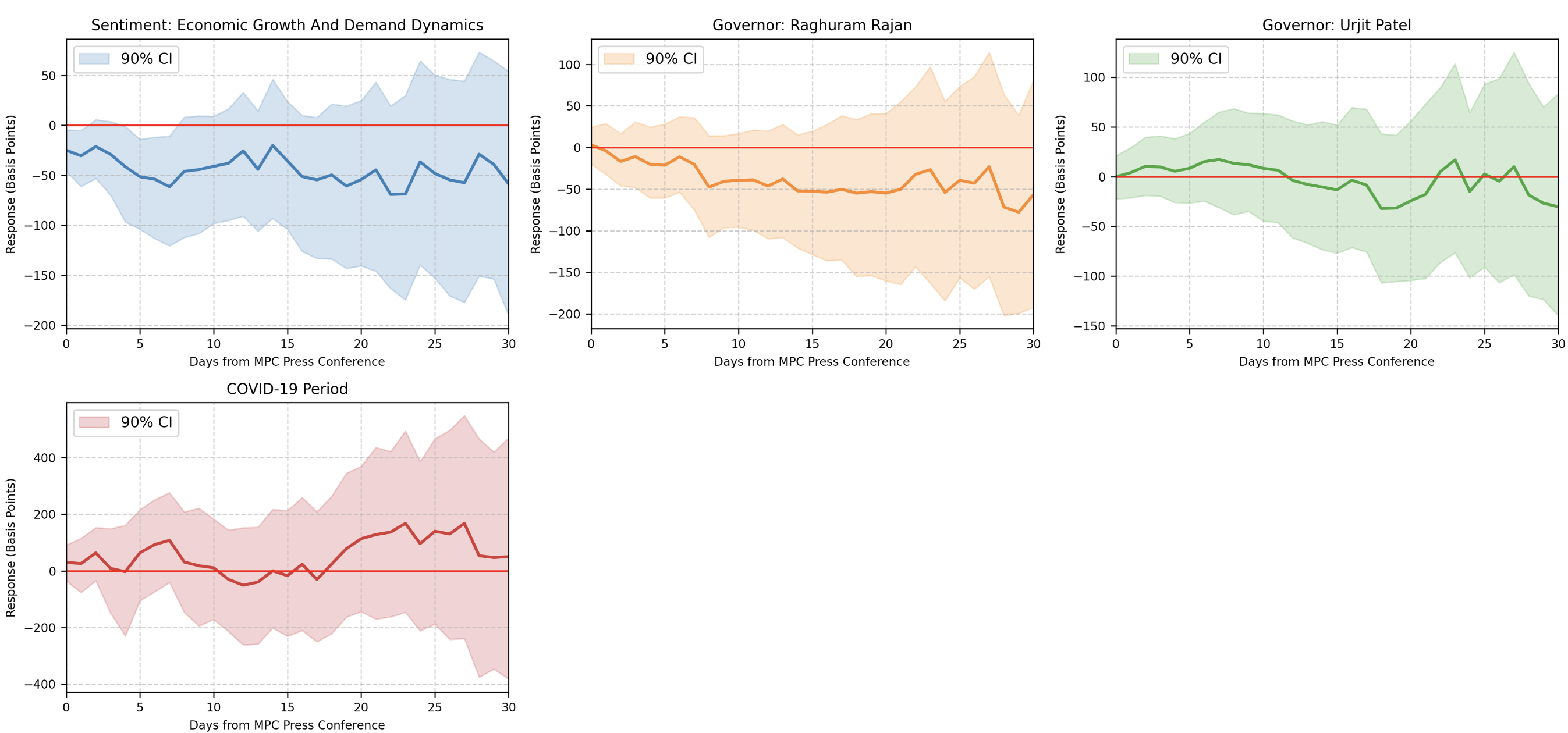}
\caption{Dynamic Responses of BSE Sensex (BSESN) to the cluster \textbf{`Economic Growth And Demand Dynamics'}}
\label{fig:economic-growth-demand}
\end{figure}
Figure \ref{fig:economic-growth-demand} illustrates the dynamic responses of the BSE Sensex to sentiment related to Economic Growth and Demand Dynamics. Surprisingly, we observe a persistent negative response of the Sensex to dovish growth sentiment, with a statistically significant decrease of up to 50 basis points over the 30-day horizon. This counter-intuitive result suggests that dovish RBI communications about growth may be interpreted as signals of potential economic overheating or future monetary tightening. Governor-specific effects reveal that this negative relationship was more pronounced during Dr. Raghuram Rajan's tenure, while Dr. Urjit Patel's period showed a mixed pattern (relative to Shri Das' period). The COVID-19 era exhibited highly volatile market responses with a tendency towards positive differential effects, though with considerable uncertainty.

\textbf{Foreign Exchange Reserves Management}
\begin{figure}
\centering
\includegraphics[width=1\linewidth]{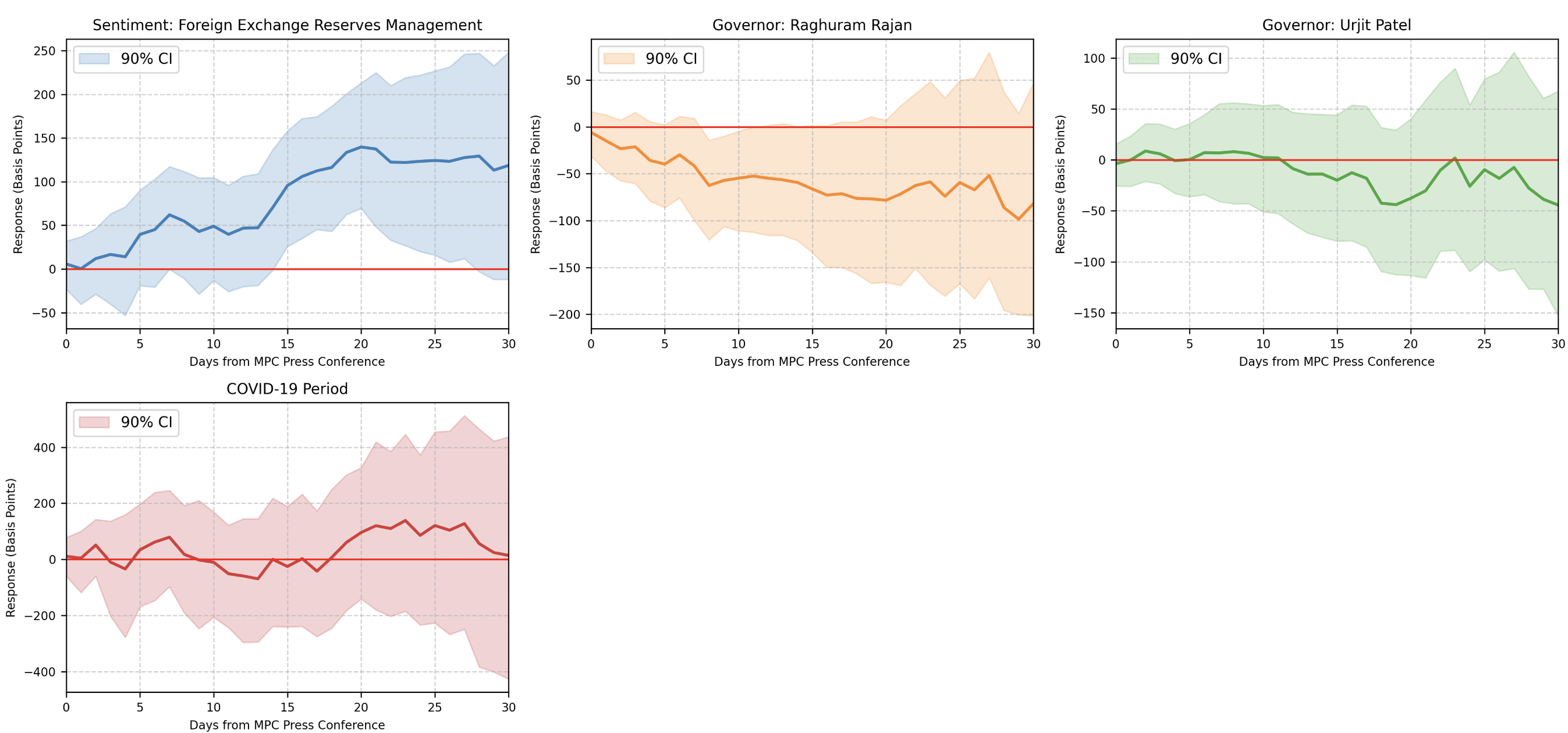}
\caption{Dynamic Responses of BSE Sensex (BSESN) to the cluster \textbf{`Foreign Exchange Reserve Management'}}
\label{fig:forex-reserve-management}
\end{figure}
Figure \ref{fig:forex-reserve-management} depicts the BSE Sensex's response to sentiment regarding Foreign Exchange Reserves Management. Notably, dovish sentiment in this domain is associated with a strong positive effect on the Sensex, becoming statistically significant after about 10 days and peaking at approximately 150 basis points around day 20. This suggests that market participants view robust foreign exchange management positively, likely interpreting it as a sign of economic stability and resilience. Interestingly, both Dr. Raghuram Rajan's and Dr. Urjit Patel's tenures show negative differential effects, indicating that the positive impact of forex reserve sentiment was less pronounced during their governorships compared to the baseline period. The COVID-19 era exhibits high volatility in responses, with a tendency towards positive effects, though marked by substantial uncertainty. The effect is immediate and persistent, reaching about 150 basis points after 30 days.  This positive reaction may indicate that markets view dovish communication about forex reserves as a sign of potential currency stability or intervention.

\textbf{Interest Rate Policy Framework}
\begin{figure}
\centering
\includegraphics[width=1\linewidth]{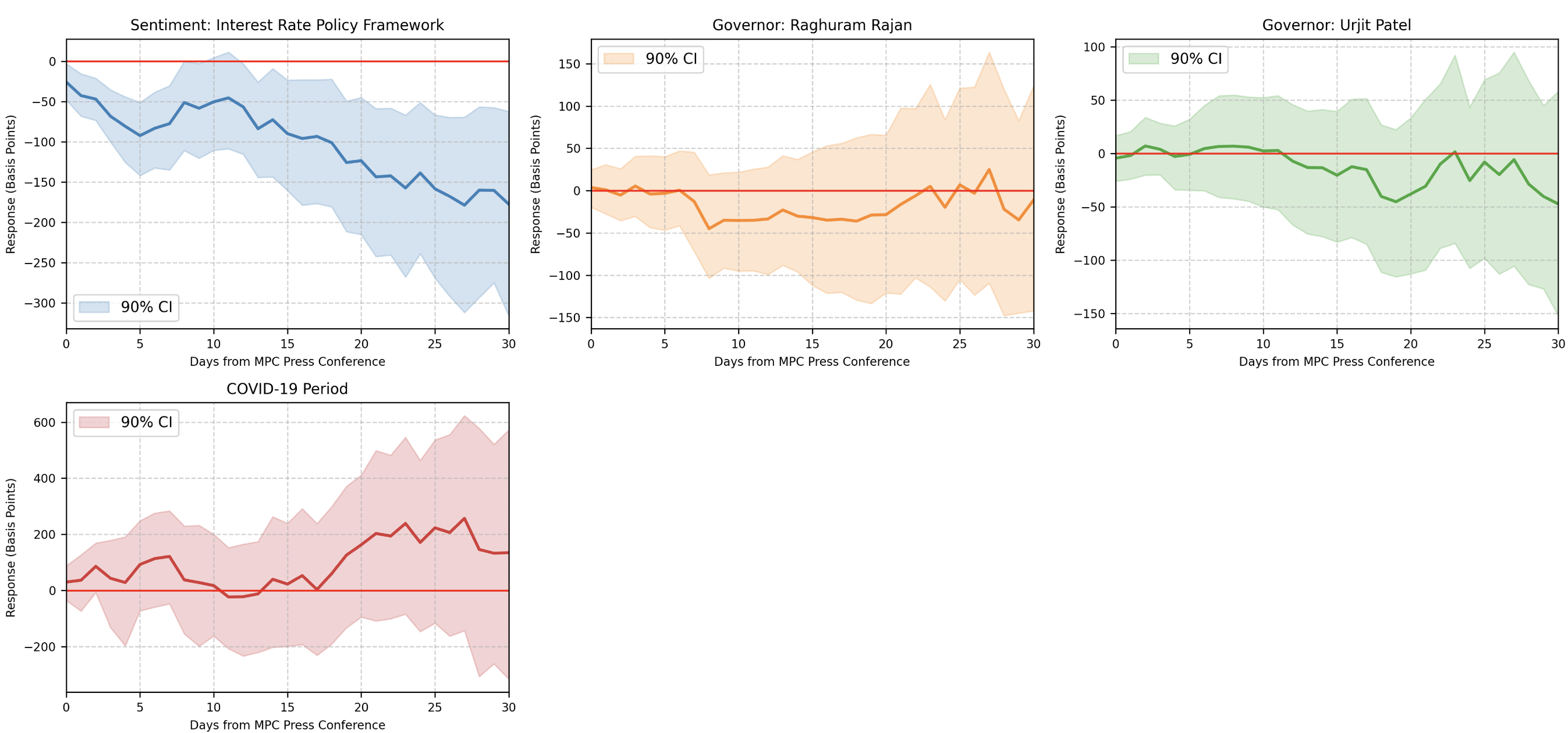}
\caption{Dynamic Responses of BSE Sensex (BSESN) to the cluster \textbf{`Interest Rate Policy Framework'}}
\label{fig:interest-rate-policy}
\end{figure}
Figure \ref{fig:interest-rate-policy} illustrates the BSE Sensex's response to sentiment related to the Interest Rate Policy Framework. The results reveal a strong and persistent negative effect, with the Sensex decreasing by up to 200 basis points over the 30-day horizon. This suggests that dovish sentiment about interest rate policies is interpreted as a signal of potential monetary tightening, leading to negative stock market reactions. Interestingly, both Dr. Raghuram Rajan's and Dr. Urjit Patel's tenures show relatively neutral differential effects, indicating consistency in market responses across governorships. The COVID-19 period exhibits a marked shift, with a strong positive differential effect, suggesting that interest rate policy communications were interpreted more favorably during the pandemic, possibly due to expectations of accommodative policies. The negative response builds over time, reaching approximately -175 basis points after 30 days.
The effect becomes statistically significant almost immediately and remains so.
This negative reaction contradicts the conventional expectation that dovish interest rate policy boosts equity markets, suggesting a potential ``information effect" (\cite{nakamura2018high}).

\textbf{Inflation Dynamics and Price Stability}
\begin{figure}
\centering
\includegraphics[width=1\linewidth]{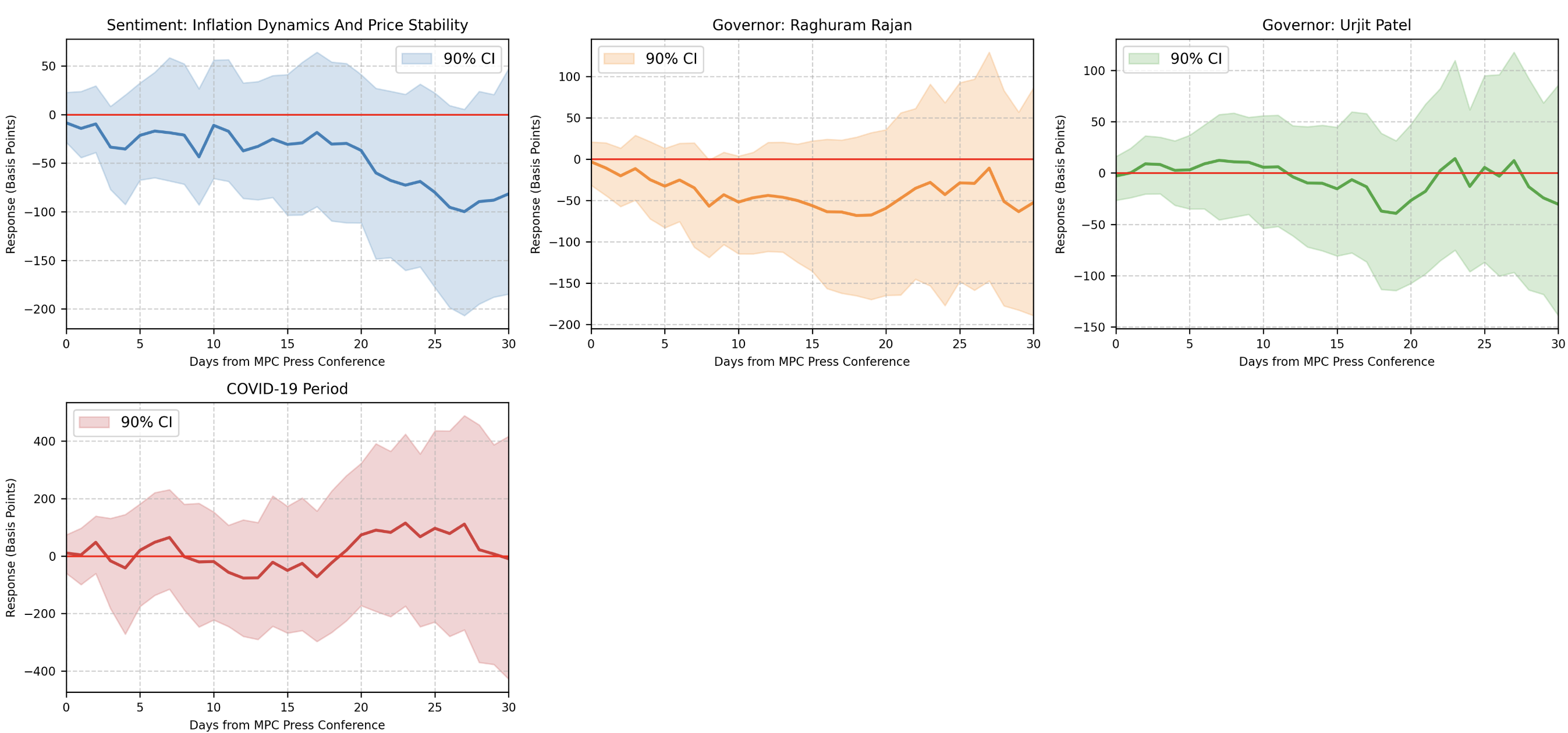}
\caption{Dynamic Responses of BSE Sensex (BSESN) to the cluster \textbf{`Inflation Dynamics Price Stability'}}
\label{fig:inflation-dynamics}
\end{figure}
 Figure \ref{fig:inflation-dynamics} illustrates the BSE Sensex's response to sentiment regarding Inflation Dynamics and Price Stability. The results show a gradually increasing negative effect, reaching approximately -100 basis points by day 30. This suggests that more dovish sentiment about inflation (i.e., less concern about inflation or more willingness to tolerate it) is associated with negative stock market reactions. This could indicate that markets interpret dovish inflation communications as a sign of potential economic overheating or future policy reversals. The responses during Dr. Raghuram Rajan's and Dr. Urjit Patel's tenures show slight negative differential effects, suggesting that dovish inflation sentiment had an even stronger negative impact during their governorships. The COVID-19 period exhibits the heightened uncertainty surrounding inflation dynamics during the pandemic. These findings highlight the complex relationship between central bank communications on inflation and market perceptions, suggesting that dovish inflation sentiment may raise concerns about long-term economic stability among market participants.

\textbf{Trade Balance and External Sector}
\begin{figure}
\centering
\includegraphics[width=1\linewidth]{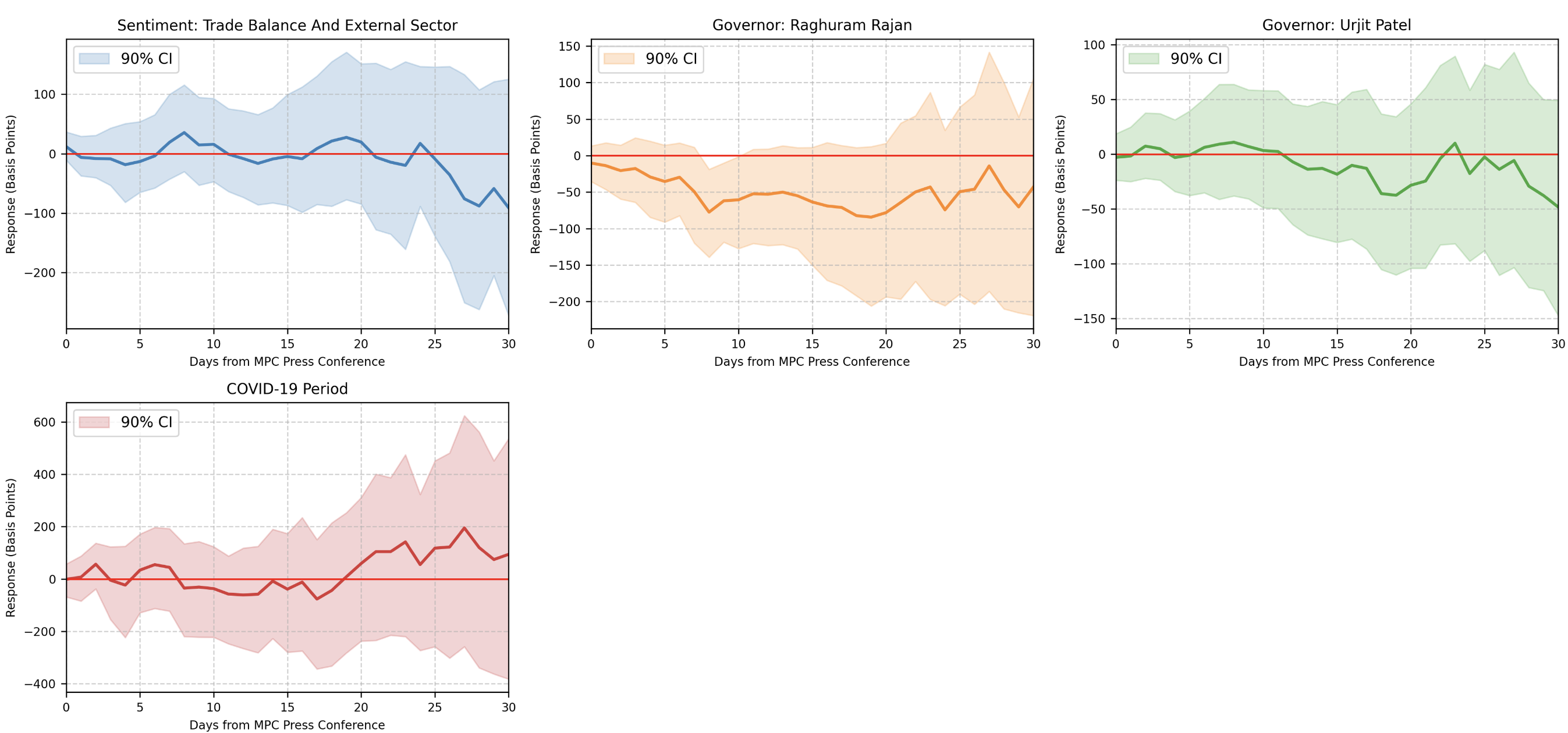}
\caption{ Dynamic Responses of BSE Sensex (BSESN) to the cluster \textbf{`Trade Balance And External Sector'}}
\label{fig:trade-balance-external}
\end{figure}
Figure \ref{fig:trade-balance-external} shows the BSE Sensex's response to sentiment regarding Trade Balance and External Sector. The results indicate a fluctuating effect over the 30-day horizon, with an initial neutral to slightly positive response turning negative after about 20 days, reaching approximately -100 basis points by day 30. This suggests that dovish sentiment about trade balance and external sector issues is initially met with market optimism, but eventually leads to concerns, possibly reflecting worries about long-term economic implications. Both Raghuram Rajan's and Urjit Patel's tenures show negative differential effects, indicating that dovish external sector communications had a more pronounced negative impact during their governorships. The COVID-19 period exhibits a marked positive differential effect with high volatility, suggesting that dovish external sector sentiment was viewed more favorably during the pandemic, possibly due to expectations of supportive policies for trade and external balance. 

\section{Conclusion}

This paper examines the effects of the Reserve Bank of India's monetary policy communications on Indian financial market over a decade, utilizing advanced natural language processing techniques for topic modeling and sentiment analysis. Our analysis reveals significant and varied impacts of sentiment across different economic topics on the stock market especially in an emerging economy context.

Notably, dovish sentiment in RBI communications generally leads to declines in equity markets, suggesting that market participants may interpret dovish language as signaling economic weakness rather than policy easing. This effect is particularly pronounced in topics related to the ``Interest Rate Policy Framework" and ``Economic Growth and Demand Dynamics," where dovish sentiment is associated with substantial negative reactions in the stock market. For instance, dovish communications about interest rates may raise concerns about future monetary tightening or economic instability, leading to a decrease in investor confidence.

Conversely, dovish sentiment regarding ``Foreign Exchange Reserves Management" has a positive impact on the stock market, indicating that investors perceive these communications as signs of economic stability and resilience. This suggests that when the RBI communicates confidence in managing foreign exchange reserves, it bolsters market optimism and supports equity valuations.

These findings underscore the importance of context-specific interpretations in central bank communications. The heterogeneous market responses across different topic clusters suggest that a one-size-fits-all communication strategy may not be optimal for central banks like the RBI. Instead, a more nuanced, topic-specific approach may be necessary to effectively manage market expectations and achieve desired policy outcomes.

Our study contributes to the literature by applying cutting-edge NLP methodologies to analyze central bank communications in an emerging market setting, providing a framework that can be extended to other economies. By integrating topic modeling with sentiment analysis, we capture the nuanced ways in which different aspects of monetary policy communications influence financial markets.

Future research could delve deeper into the mechanisms behind the differential market responses to topic-specific sentiments. Expanding the analysis to include a broader range of financial indicators and real economic variables could offer a more comprehensive understanding of the transmission channels of monetary policy communication. Comparative studies with other emerging market central banks might also help identify common patterns and unique features in communication effectiveness across different institutional contexts.

Understanding the complex interactions between central bank communications and financial markets is crucial for effective monetary policy implementation and financial stability. As central banks continue to rely on communication as a key policy tool, especially in emerging markets, tailoring strategies to account for how markets interpret different economic topics becomes increasingly important.

\bibliography{references}

\newpage

\begin{center}
\textbf{ONLINE B.3 TO}

\vspace{0.5cm}

\textbf{Words that Move Markets: Quantifying the Impact of RBI's Monetary Policy Communications on Indian Financial Markets} \newline
\textbf{A Step-by-Step Approach}

\vspace{0.5cm}

\end{center}

\section*{A \hspace{0.2cm} Algorithm for Sentence-Level Text Processing}

The following algorithm describes the sequential steps used for processing central bank communications at the sentence level, as outlined in the main text.

\begin{enumerate}
    \item \textbf{Data Input and Consolidation}
    \begin{enumerate}
        \item Read multiple sheets from an Excel file using \texttt{pandas}.
        \item Combine data from all sheets into a single \texttt{DataFrame}.
        \item Assign unique paragraph IDs and append metadata (e.g., source, date).
    \end{enumerate}
    \item \textbf{Sentence Tokenization}
    \begin{enumerate}
        \item Use NLTK's \texttt{sent\_tokenize} function to split paragraphs into sentences.
        \item Assign metadata (source, date, paragraph ID) to each sentence.
    \end{enumerate}
    \item \textbf{Irrelevant Sentence Filtering}
    \begin{enumerate}
        \item Define a minimum word count threshold (default: 3 words) (figure 8).
        \item  Remove sentences with fewer words than the threshold.
        \item Purpose: Eliminate non-substantive content.
    \end{enumerate}
    \item \textbf{Text Statistics Generation}
    \begin{enumerate}
        \item  Calculate metrics for both unfiltered and filtered data:
        \begin{enumerate}
            \item Number of paragraphs
            \item Number of sentences
            \item Total word count
            \item Average sentence length
        \end{enumerate}
    \end{enumerate}
    \item \textbf{Sentence Length Distribution Analysis}
    \begin{enumerate}
        \item Plot the distribution of sentence lengths.
        \item Calculate and overlay statistics (mean, median, mode) on the plot. 
        \item Purpose: Visualize sentence length patterns in the corpus.
    \end{enumerate}
    \item \textbf{Threshold Selection and Validation}
    \begin{enumerate}
        \item Calculate a suggested threshold based on distribution statistics.
        \item  Sample and print example sentences near the threshold. 
        \item Allow for manual review and adjustment of the threshold.
    \end{enumerate}
    \item \textbf{Integrated Processing Pipeline}
    \begin{enumerate}
        \item Combine all steps into a single function (\texttt{process\_data}).
        \item Execute the entire workflow from data input to final filtering.
        \item Generate comparative metrics for filtered and unfiltered datasets.
    \end{enumerate}
    \item \textbf{Data Output}
    \begin{enumerate}
        \item Prepare the processed, sentence-level data for further analysis.
        \item Retain all relevant metadata with each sentence.
    \end{enumerate}
\end{enumerate}

\begin{figure}
        \centering
        \includegraphics[width=0.99\linewidth]{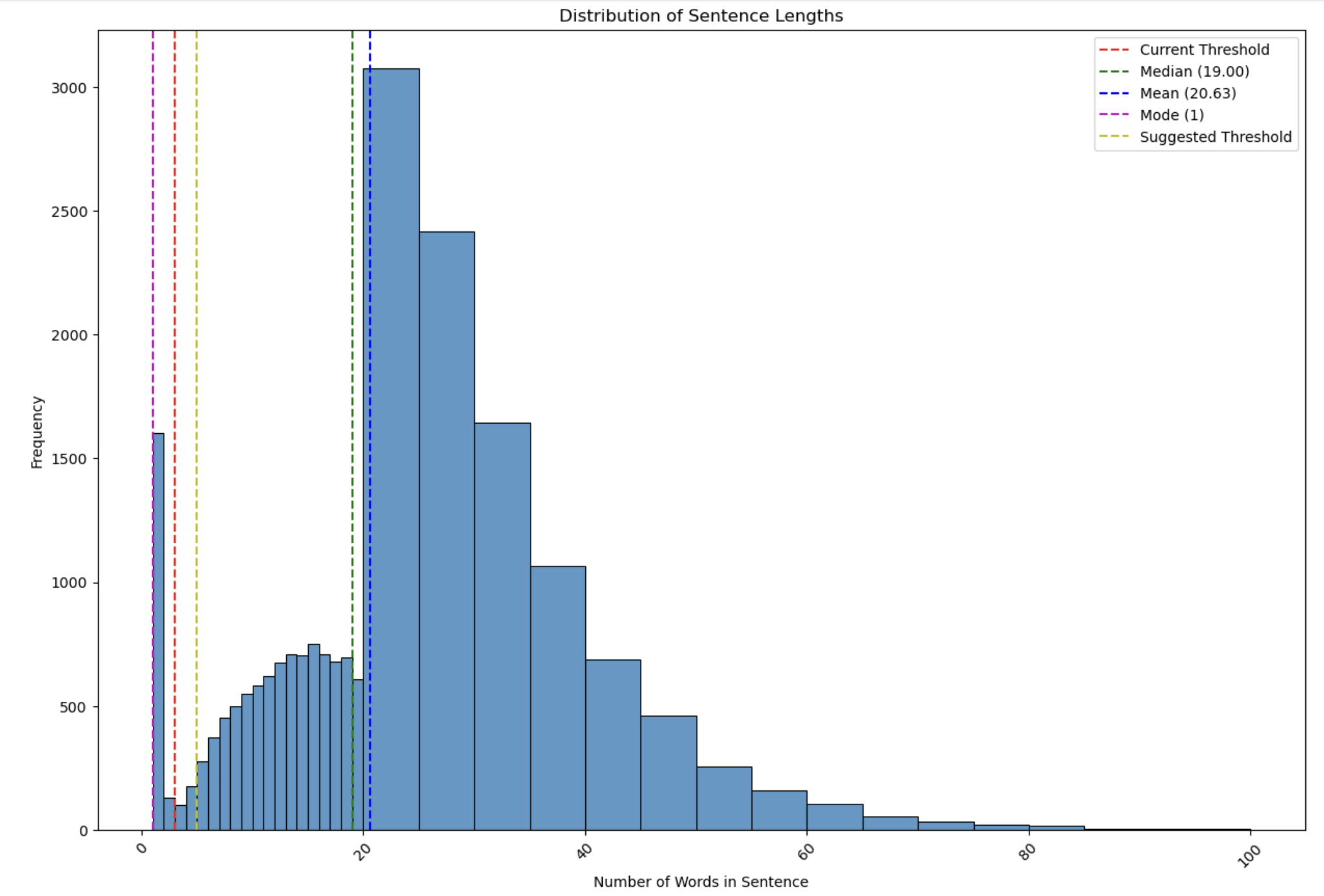}
        \caption{Selecting Threshold}
        \label{fig:threshold-histogram}
\end{figure}

\section*{B\quad Algorithm for Topic Representation}

\vspace{0.5cm}

\begin{figure}
    \centering
    \includegraphics[width=0.7\linewidth]{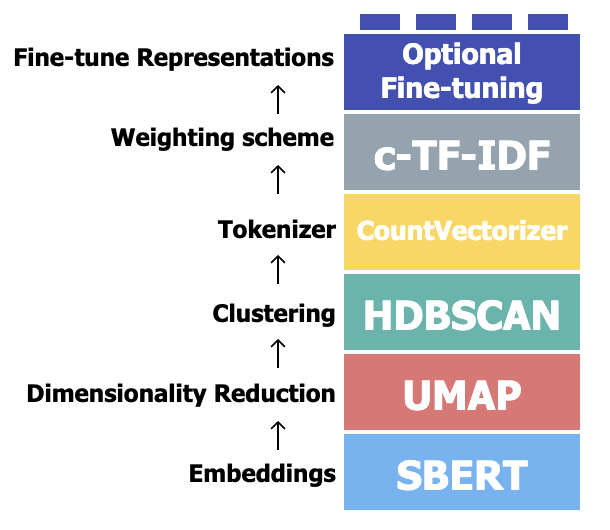}
    \caption{BERTopic algorithm process (Source: \cite{grootendorst2024bertopic}}
    \label{fig:bert}
\end{figure}

\subsection*{B.1 \hspace{0.3cm} Understanding TF-IDF and c-TF-IDF in BERTopic}

The below explanation describes TF-IDF (Term Frequency-Inverse Document Frequency) and its variant c-TF-IDF (class-based TF-IDF), key components of the BERTopic algorithm used in our analysis, as described in Section 3 of the main text.

\begin{enumerate}
    \item \textbf{TF-IDF}: A method to evaluate word importance in a document within a collection of documents.
    \begin{enumerate}
        \item \textbf{TF (Term Frequency)}: How often a word appears in a document.
        \item \textbf{IDF (Inverse Document Frequency)}: How unique or rare the word is across all documents.
        \item \textbf{TF-IDF score} = TF * IDF. Higher scores indicate words that are both frequent in a document and unique across documents.
    \end{enumerate}
    
    \item \textbf{c-TF-IDF}: A variation of TF-IDF adapted for topic modeling in BERTopic.
    \begin{enumerate}
        \item Treats each topic (cluster of documents) as a single document.
        \item Calculates word frequency within each topic.
        \item Determines word importance across all topics.
    \end{enumerate}
    
    \item The c-TF-IDF score for a word in a topic is calculated as follows:
    \begin{enumerate}
        \item \textbf{TF}: The frequency of the word in the topic, normalized by total words in that topic.
        \item \textbf{IDF}: The logarithm of total topics divided by the number of topics containing the word.
        \item \textbf{Final score}: TF * IDF.
    \end{enumerate}
    
    \item \textbf{Key differences from traditional TF-IDF}:
    \begin{enumerate}
        \item Uses topics instead of individual documents.
        \item Identifies words uniquely important to specific topics.
        \item Provides better topic representation in the context of the entire corpus.
    \end{enumerate}
    
    \item \textbf{Advantages in topic modeling}:
    \begin{enumerate}
        \item Captures topic-specific terminology more effectively.
        \item Reduces the impact of common words across all topics.
        \item Offers a more nuanced representation of topics compared to simple word frequency.
    \end{enumerate}
\end{enumerate}

\subsection*{B.2 \hspace{0.2cm} Explanation of Transformer Models and the \texttt{"all-MiniLM-L6-v2"} Embedding Model}

The below explanation describes Transformer models and the specific \texttt{"all-MiniLM-L6-v2"} embedding (SentenceTransform) model used in our analysis, as described in Section 3 of the main text.

\begin{enumerate}
    \item \textbf{Introduction to Transformer models}:
    \begin{enumerate}
        \item Advanced AI models designed to understand and process human language.
        \item Use a mechanism called ``attention" to focus on important parts of text.
        \item Can understand context and nuances in language better than previous models.
        \item Form the basis for many modern natural language processing tasks.
    \end{enumerate}

    \item \textbf{Overview of \texttt{"all-MiniLM-L6-v2"}}:
    \begin{enumerate}
        \item A compact Transformer-based model developed by Microsoft and the University of Mannheim.
        \item Designed to be smaller and faster than larger models like BERT, while maintaining high performance.
        \item Trained on a diverse corpus of web content and fine-tuned for semantic similarity tasks.
    \end{enumerate}

    \item \textbf{Key features relevant to financial language analysis}:
    \begin{enumerate}
        \item Trained on diverse text sources, including financial and economic content.
        \item Optimized for sentence-level semantic similarity, crucial for comparing financial statements.
        \item Supports multiple languages, beneficial for analyzing global financial communications.
    \end{enumerate}

    \item \textbf{Advantages for capturing financial language nuances}:
    \begin{enumerate}
        \item High performance on semantic textual similarity benchmarks, allowing detection of subtle differences in economic statements.
        \item Effective at handling domain-specific vocabulary common in central bank communications.
        \item Capable of understanding context-dependent meanings of financial terms.
    \end{enumerate}

    \item \textbf{Technical specifications}:
    \begin{enumerate}
        \item Model size: 80MB (compared to several GB for larger models).
        \item Embedding dimension: 384 (balanced between expressiveness and efficiency).
        \item Vocabulary size: 30,522 tokens (including financial and economic terms).
    \end{enumerate}

    \item \textbf{Comparative advantages}:
    \begin{enumerate}
        \item Faster processing time compared to larger models, suitable for analyzing large volumes of policy documents.
        \item Lower computational resource requirements, making it accessible for various research setups.
        \item Performance comparable to much larger models on financial text tasks.
    \end{enumerate}
\end{enumerate}

\subsection*{B.3 \hspace{0.3cm} Explanation of Dimensionality Reduction Techniques: PCA, t-SNE, and UMAP}

The below explanation describes three common dimensionality reduction techniques and their relevance to our analysis, as mentioned in Section 3 of the main text.

\begin{enumerate}
    \item \textbf{Principal Component Analysis (PCA)}:
    \begin{enumerate}
        \item A linear dimensionality reduction technique.
        \item Identifies the directions (principal components) of maximum variance in the data.
        \item Projects data onto these components, preserving global structure.
        \item Computationally efficient but may miss complex, non-linear relationships.
    \end{enumerate}

    \item \textbf{t-Distributed Stochastic Neighbor Embedding (t-SNE)}:
    \begin{enumerate}
        \item A non-linear technique focused on preserving local structure.
        \item Particularly good at visualizing high-dimensional data in 2D or 3D space.
        \item Emphasizes the preservation of similarities between nearby points.
        \item Can be computationally intensive for large datasets.
    \end{enumerate}

    \item \textbf{Uniform Manifold Approximation and Projection (UMAP)}:
    \begin{enumerate}
        \item A non-linear technique that balances preservation of local and global structure.
        \item Based on concepts from topological data analysis and manifold learning.
        \item Often faster than t-SNE, especially for larger datasets.
        \item Can be used for general dimensionality reduction, not just visualization.
    \end{enumerate}

    \item \textbf{Comparison in the context of topic modeling}:
    \begin{enumerate}
        \item \textbf{PCA}: Good for initial dimensionality reduction, but may miss nuanced topic structures.
        \item \textbf{t-SNE}: Excellent for visualizing topic clusters, but may distort global relationships.
        \item \textbf{UMAP}: Balances local and global structure, suitable for both visualization and further analysis.
    \end{enumerate}

    \item \textbf{Advantages of UMAP for our analysis}:
    \begin{enumerate}
        \item Preserves both local and global structure of the high-dimensional space.
        \item Enables identification of coherent economic topics with complex interrelationships.
        \item Computationally efficient for large datasets of policy communications.
        \item Provides a good foundation for subsequent clustering steps in topic modeling.
    \end{enumerate}
\end{enumerate}

\subsection*{B.4 \hspace{0.3cm} Understanding Topic Representation in BERTopic}

\vspace{0.5cm}

The below algorithm describes the sequential steps BERTopic employs to create topic representations from text data, as described in the main text.

\begin{enumerate}[leftmargin=*]
    \item Start with embeddings. Apply SBERT (Sentence-BERT) to generate embeddings for each document in the corpus.
    \item Perform dimensionality reduction. Use UMAP to reduce the high-dimensional embeddings to a more manageable space.
    \begin{enumerate}
        \item We set n\_neighbors=15 and n\_components=5. The n\_neighbors parameter determines the local neighborhood size for mapping, while n\_components specifies the dimensionality of the reduced space. 
        \item These settings balance local structure preservation with effective global   dimensionality reduction.
    \end{enumerate}
  
    \item IF the dimensionality reduction significantly alters the original embedding space:
    \begin{enumerate}
        \item Consider adjusting UMAP parameters OR
        \item Explore alternative dimension reduction techniques
    \end{enumerate}
    
    \item Apply clustering. Utilize HDBSCAN to cluster the reduced embeddings into topics.

    \begin{enumerate}
        \item We set min\_cluster\_size=150, ensuring each identified topic has substantial support in the data, thus minimizing spurious topics. 
        \item This parameter represents the minimum number of data points required to constitute a topic, focusing the analysis on significant themes in the communications.
    \end{enumerate}
    
    \item IF the clustering results are unsatisfactory:
    \begin{enumerate}
        \item Adjust HDBSCAN parameters OR
        \item Consider alternative clustering algorithms
    \end{enumerate}
    
    \item Create bag-of-words representations. Use CountVectorizer to convert documents within each cluster into token count matrices.
    
    \item Apply c-TF-IDF weighting. Calculate class-based TF-IDF scores to identify important words for each topic.
    
    \item Generate topic representations. Select top n words based on c-TF-IDF scores to represent each topic.
    \begin{enumerate}
        \item IF topic representations need refinement:
        \begin{enumerate}
            \item Adjust the number of top words OR
            \item Implement additional filtering techniques
        \end{enumerate}
    \end{enumerate}
 
    \item Perform topic reduction and merging (optional). Combine similar topics based on semantic similarity or other criteria.
    
    \item Fine-tune topic representations (optional). Apply techniques such as guided topic modeling or hierarchical topic modeling for further refinement.
\end{enumerate}

\section*{ C\quad Examples of Sentiment Classification}

To illustrate how the pre-trained RoBERTa model (on FOMC data) classifies statements, we provide the following examples along with their classifications, reasoning, and potential market implications: \newline

1. Statement: ``Such a directive would imply that any tightening should be implemented promptly if developments were perceived as pointing to rising inflation."

   Classification: Hawkish (Score: -0.75)

   Reasoning: The phrase ``tightening should be implemented promptly" signals a readiness to raise interest rates or reduce liquidity in response to inflation concerns.

   Market Implication: Bond yields may rise, and equity markets might react negatively due to expectations of tighter financial conditions. \newline

2. Statement: ``The International Monetary Fund projects that global economic growth in 2019 will be the slowest since the financial crisis."

   Classification: Dovish (Score: 0.62)

   Reasoning: This statement reflects concerns over slow global growth, suggesting a potential need for accommodative policies.

   Market Implication: Bond yields may fall as expectations for rate cuts increase, potentially supporting equity markets. \newline

3. Statement: ``The volume of world trade can shrink by 13-32 per cent in 2020, as projected by the World Trade Organisation (WTO)."

   Classification: Dovish (Score: 0.58)

   Reasoning: The statement indicates significant downside risks to global trade, which might necessitate accommodative measures.

   Market Implication: Bond yields may decline, while equity markets might have mixed reactions depending on the expected severity of trade contraction and policy response. \newline

4. Statement: ``World services trade deteriorated in the first quarter of 2020 due to a broad-based loss of momentum in passenger air travel, container shipping, financial and ICT services."

   Classification: Neutral to Slightly Dovish (Score: 0.15)

   Reasoning: While highlighting a negative trend, it doesn't explicitly signal an immediate policy stance. The slight dovish tone comes from the implication that easing measures might be considered if the slowdown persists.

   Market Implication: Markets might not react strongly to this alone, but it could reinforce expectations of easing in the context of broader dovish signals. \newline

These examples demonstrate how our model captures nuances in monetary policy communication, assigning scores that reflect both the direction and intensity of the sentiment. This granular approach allows for a more precise examination of how linguistic shifts in central bank communication may influence financial markets.

\newpage

\section*{ D\quad Topic Structure with Representative Documents}

\begin{xltabular}{\textwidth}{|C{1cm}|C{4cm}|L{10cm}|}
\caption{Detailed Topic Structure with Representative Documents} \label{tab:detailed_topics_new} \\
\toprule
\textbf{Topic} & \textbf{Custom Name} & \textbf{Representative Docs} \\ 
\midrule
\endfirsthead

\multicolumn{3}{c}%
{{\bfseries \tablename\ \thetable{} -- continued from previous page}} \\
\toprule
\textbf{Topic} & \textbf{Custom Name} & \textbf{Representative Docs} \\ 
\midrule
\endhead

\midrule
\multicolumn{3}{r}{{Continued on next page}} \\
\midrule
\endfoot

\bottomrule
\endlastfoot

1 & Inflation Dynamics and Price Stability & \begin{itemize}[noitemsep, topsep=0pt]
    \item We analyse the headline inflation and of course, as a part of that other aspects like food inflation, non-food inflation, non-oil inflation, and fuel inflation; all these things are analysed.
    \item Retail inflation, measured by the year-on-year change in the CPI, rose from 4.9 per cent in May to 5 per cent in June, driven by an uptick in inflation in fuel and in items other than food and fuel even as food inflation remained muted due to lower than usual seasonal uptick in prices of fruits and vegetables in summer months.
    \item We analyse the headline inflation and of course, as a part of that other aspects like food inflation, non-food inflation, non-oil inflation, and fuel inflation; all these things are analysed.
\end{itemize} \\ \midrule

2 & Trade Balance and External Sector & \begin{itemize}[noitemsep, topsep=0pt]
    \item The contraction in merchandise imports outpaced that of merchandise exports in April, resulting in a narrowing of the trade deficit.
    \item With services exports moderating, the widening of the merchandise trade deficit could lead to a modest increase in the current account deficit (CAD) during Q2.
    \item During the first three quarters of 2023-24, India’s current account deficit (CAD) narrowed significantly on account of a moderation in merchandise trade deficit coupled with robust growth in services exports and strong remittances.
\end{itemize} \\ \midrule

3 & Economic Growth and Demand Dynamics & \begin{itemize}[noitemsep, topsep=0pt]
    \item Taking all these factors into consideration, real GDP growth for 2023-24 is projected at 6.5 per cent with Q2 at 6.5 per cent; Q3 at 6.0 per cent; and Q4 at 5.7 per cent.
    \item Taking all these factors into consideration, real GDP growth for 2024-25 is projected at 7.0 per cent with Q1 at 7.2 per cent; Q2 at 6.8 per cent; Q3 at 7.0 per cent; and Q4 at 6.9 per cent.
    \item Growth in steel consumption, cement output, and production and imports of capital goods suggest continued buoyancy in investment activity.
\end{itemize} \\ \midrule

4 & Foreign Exchange Reserves Management & \begin{itemize}[noitemsep, topsep=0pt]
    \item In 2020-21 (April-July), net foreign portfolio investment (FPI) in equities at US\$ 5.3 billion was higher than US\$ 1.2 billion a year ago.
    \item Net foreign portfolio investment was of the order of US\$ 8.6 billion in 2019-20 (up to February 4) as against net outflows of US\$ 14.2 billion in the same period last year.
    \item On the financing side, net foreign direct investment inflows picked up in March 2020 to US\$ 2.9 billion from US\$ 0.8 billion a year ago.
\end{itemize} \\ \midrule

5 & Foreign Investment in Securities Markets & \begin{itemize}[noitemsep, topsep=0pt]
    \item Tripartite Repo in Government Securities Market: The Working Group on Enhancing Liquidity in the Government Securities and Interest Rate Derivatives (Chairman: Shri R. Gandhi) had recommended introduction of tripartite repo to develop a term repo market.
    \item The investment limit in Government securities by foreign portfolio investors (FPIs), registered with the Securities and Exchange Board of India (SEBI) is currently capped at USD 30 billion of which USD 5 billion is reserved for long term investors.
    \item With a view to improving the depth and liquidity in the domestic foreign exchange market, it has been decided to allow foreign portfolio investors to participate in the domestic exchange traded currency derivatives market to the extent of their underlying exposures plus an additional US\$ 10 million.
\end{itemize} \\ \midrule

6 & Interest Rate Policy Framework & \begin{itemize}[noitemsep, topsep=0pt]
    \item Today, we have done a 25 basis points cut in the policy rate.
    \item Monetary Policy transmission encompasses the whole continuum of interest rates, of course the central bank only determines the overnight policy rate.
    \item Earlier we cut rates but there was not much transmission, my belief is both with the MCLR and its effects which we are already seeing in the market, as I said this morning, we are already seeing a 25 basis points median rate reduction and a 50 basis points overnight reduction even before the policy rate cut.
\end{itemize} \\ \midrule

7 & Financial Markets and Volatility & \begin{itemize}[noitemsep, topsep=0pt]
    \item I will never predict what financial markets do.
    \item These developments have led to heightened volatility in global financial markets as reflected in sizeable two-way movements in bond yields, fall in equity markets and the US dollar shedding its gains from its peak of September 2022.
    \item Financial markets are seeing excessive volatility.
\end{itemize} \\ \midrule

8 & Banking Sector Credit Dynamics & \begin{itemize}[noitemsep, topsep=0pt]
    \item Finally, the slowdown in credit growth is more pronounced in public sector banks, but how much of this is because of needed bank balance sheet restructuring, repayments of stressed loans, or increased risk aversion is to be established.
    \item It was to give banks more flexibility on their balance sheet going forward as some of the demands on the banks pick up, as credit growth picks up.
    \item So, banks are not inconvenienced by it.
\end{itemize} \\ \bottomrule

\end{xltabular}

\section*{E\quad Baseline Responses with Robustness Check}

In order to establish the unconditional relationship between sentiment and market returns, we estimate the following simplest version of the Local Projections Model:

$$y_{i,t+h} - y_{i,t} = \beta_0^{(h)} + \beta_1^{(h)} S_{j,t} + \varepsilon_{i,t+h}$$

\begin{figure}[htbp]
    \centering
    \begin{subfigure}{0.45\linewidth}
        \centering
        \includegraphics[width=\linewidth]{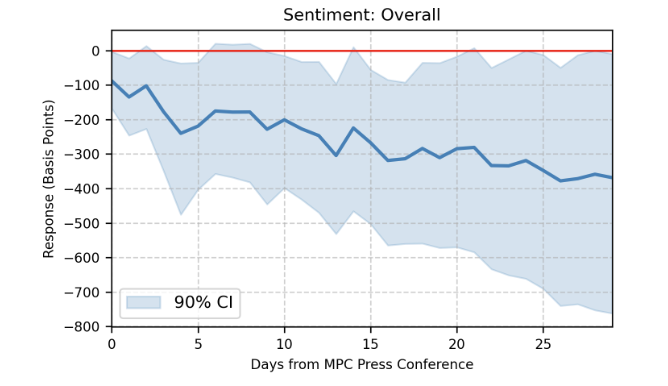}
        \label{fig:sole-regressor}
    \end{subfigure}%
    \hfill
    \begin{subfigure}{0.45\linewidth}
        \centering
        \includegraphics[width=\linewidth]{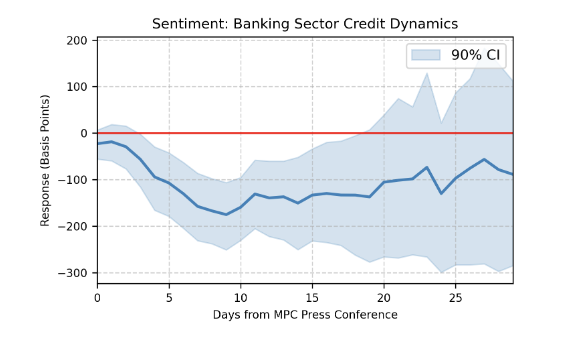}
        \label{fig:sole-regressor-bankdynamics}
    \end{subfigure}

    \vspace{1em}

    \begin{subfigure}{0.45\linewidth}
        \centering
        \includegraphics[width=\linewidth]{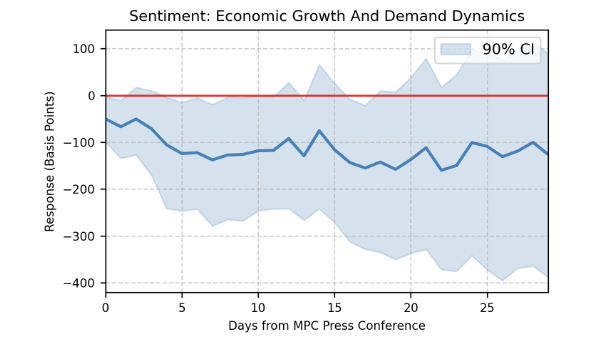}
        \label{fig:sole-regressor-ecogr}
    \end{subfigure}%
    \hfill
    \begin{subfigure}{0.45\linewidth}
        \centering
        \includegraphics[width=\linewidth]{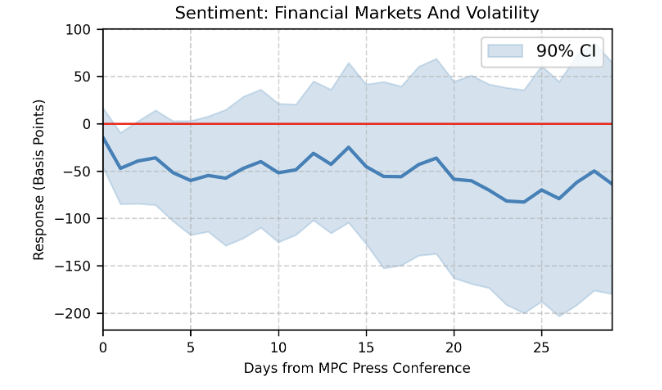}
        \label{fig:sole-regressor-fmv}
    \end{subfigure}

    \vspace{1em}

    \begin{subfigure}{0.45\linewidth}
        \centering
        \includegraphics[width=\linewidth]{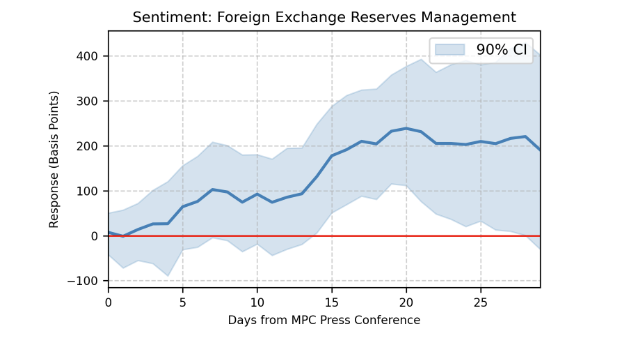}
        \label{fig:sole-reg-forex}
    \end{subfigure}%
    \hfill
    \begin{subfigure}{0.45\linewidth}
        \centering
        \includegraphics[width=\linewidth]{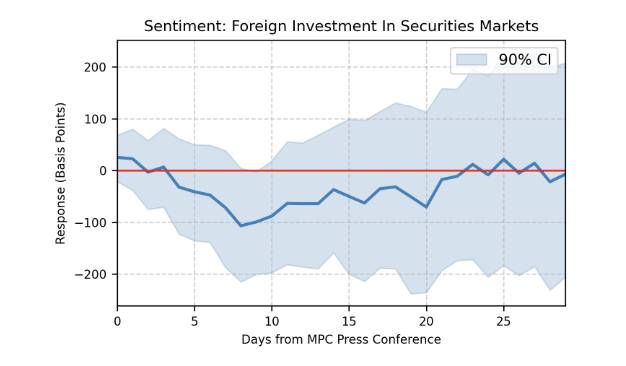}
        \label{fig:sole-regressor-foreignsec}
    \end{subfigure}

    \vspace{1em}

    \begin{subfigure}{0.45\linewidth}
        \centering
        \includegraphics[width=\linewidth]{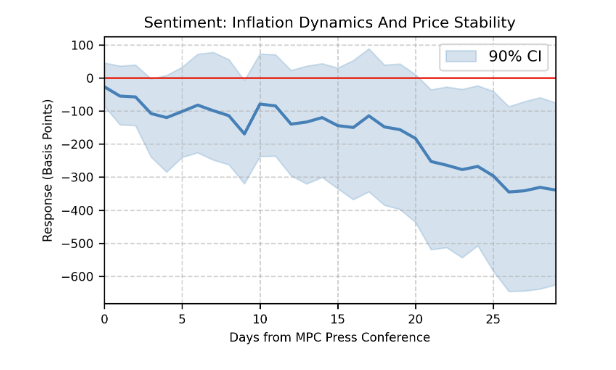}
        \label{fig:sole-reg-inflation}
    \end{subfigure}%
    \hfill
    \begin{subfigure}{0.45\linewidth}
        \centering
        \includegraphics[width=\linewidth]{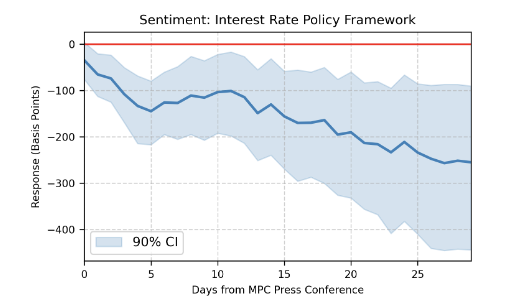}
        \label{fig:sole-reg-interest}
    \end{subfigure}

    \caption{Baseline Local Projections Model with Sole Regressor: Sentiment}
    \label{fig:juxtaposed-images}
\end{figure}

Baseline Local Projections Model with Governor and COVID Dummies along with Monetary Policy Shock series (\cite{lakdawala2024measuring}):

$$y_{i,t+h} - y_{i,t} = \beta_0^{(h)} + \beta_1^{(h)} S_{j,t}  + \beta_2^{(h)} \text{COVID}_{t} + \beta_3^{(h)} \text{Rajan}_{t} + \beta_4^{(h)} \text{Patel}_{t} + \beta_5^{(h)} X_{t} + \varepsilon_{i,t+h}$$

\begin{figure}
    \centering
    \includegraphics[width=1\linewidth]{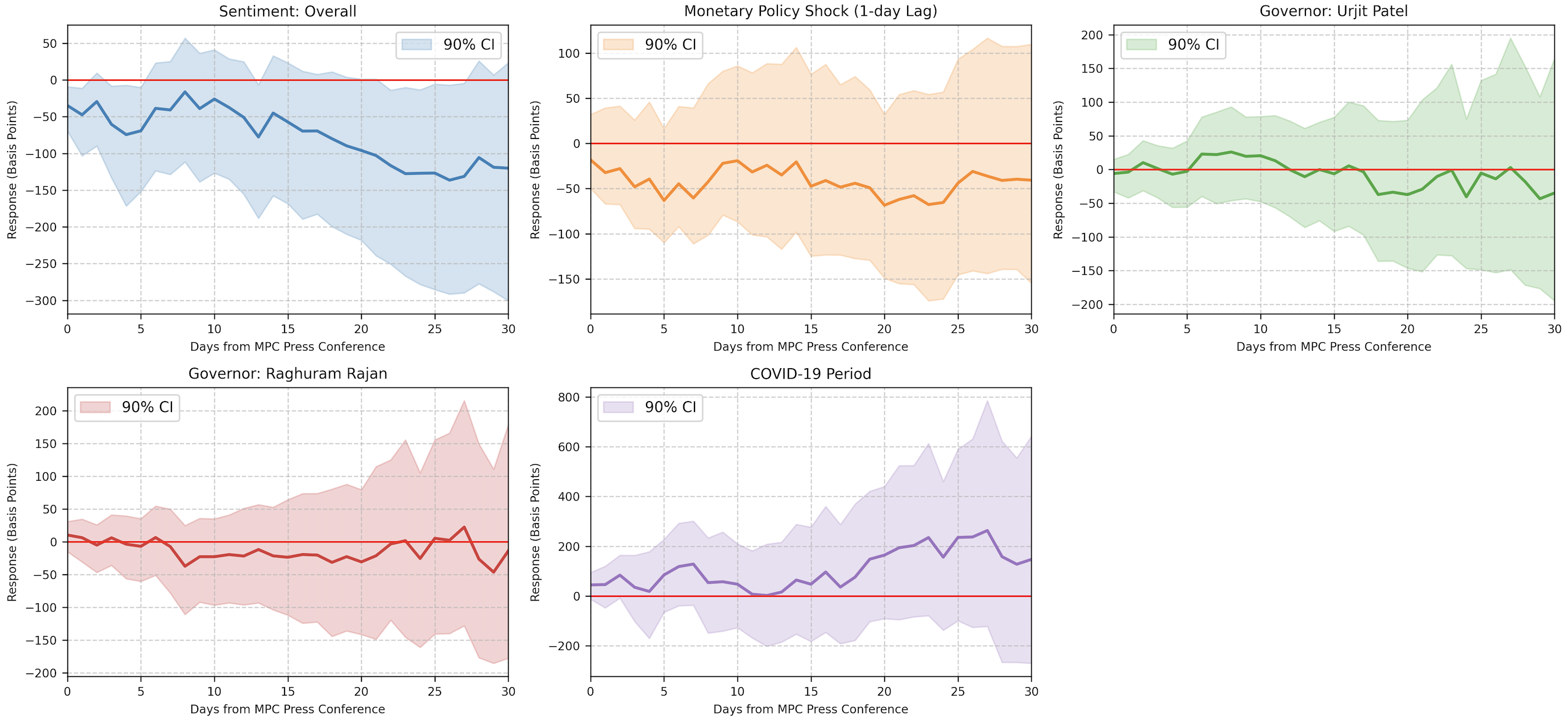}
    \caption{Additional Control Variable}
    \label{fig:Feb23-Gov-COVID-MPshock}
\end{figure}

However, to generate the impulse responses presented in Figure \ref{fig:Feb23-Gov-COVID-MPshock} onward using this specification, we were unable to utilize all the available RBI communication data. This limitation was necessary to align our analysis with the availability of the Monetary Policy (MP) Shock series, which is specifically available uptil February 2023.

Further robustness checks with sentiment analysis at cluster level, revealed considerably significant results especially for `Banking Sector Credit Dynamics', `Foreign Exchange Reserves Management' and `Interest Rate Policy Framework'. We also obtained significant results with clear trend when we included other control variables such as Inflation Growth Rate and Economic Policy Uncertainty Index (\cite{baker2024economic}).

\end{document}